\documentclass[pre,nofootinbib]{revtex4-1}
\usepackage{graphicx}
\usepackage{amssymb,amsfonts,amsthm,bm}

\newcommand{\beq}[0]{\begin{equation}}
\newcommand{\eeq}[0]{\end{equation}}

\newcommand{\thet}{\vartheta}

\newcommand{\ds}{\displaystyle}

\newcommand{\pa}{\partial}

\newcommand{\lesq}{\leqslant}

\begin{document}

\author{V. Koukouloyannis}
\affiliation{Department of Physics, University of Thessaloniki, GR-54124 Thessaloniki, Greece}
\author{G. Voyatzis}
\affiliation {Department of Physics, University of Thessaloniki, GR-54124 Thessaloniki, Greece}
\author{P.G. Kevrekidis}
\affiliation{Department of Mathematics and Statistics, University of
Massachusetts, Amherst MA 01003-4515}

\title{Dynamics of Three Non-co-rotating Vortices in Bose-Einstein Condensates}

\begin{abstract}
In this work we use standard Hamiltonian-system techniques in order to study the dynamics of three vortices with alternating charges in a confined Bose-Einstein condensate. In addition to being motivated by recent experiments, this
system offers a natural vehicle for the exploration of the transition of
the vortex dynamics from ordered to progressively chaotic behavior.
In particular, it  possesses two integrals of motion, the {\it energy} (which is expressed through the Hamiltonian $H$) and the 
{\it angular momentum}  $L$ of the system. By using the integral of the angular momentum, we reduce the system to a two degree-of-freedom one with $L$ as a parameter and reveal the topology of the phase space through the method of Poincar\'e surfaces of section. 

We categorize the various motions that appear in the different regions of the sections and we study the major bifurcations that occur to the families of periodic motions of the system. Finally, we correspond the orbits on the surfaces of section to the real space  motion of the vortices in the plane.
\end{abstract}

\maketitle

\section{Introduction}

The exploration of 2D point vortex dynamics is a fascinating topic
with a rich history for over a century, starting, arguably, with 
the fundamental contribution of Lord Kelvin~\cite{kel1} and gradually 
progressing
to the consideration of higher numbers of vortices
(see e.g.~\cite{havel,ziff}) and of not only symmetric but also of asymmetric
equilibria thereof (see e.g.~\cite{aref0}). This large volume of relevant
fluid literature extending from few vortex clusters to large
scale vortex crystals has been summarized in numerous publications; see
e.g. the review~\cite{aref1} and the book~\cite{newton1}. Along 
the way, this effort has also triggered experimental investigations
considering not only stable vortex patterns in rotating superfluid
$^{4}$He~\cite{yarmchuk}, but also electron columns confined in 
Malmberg-Penning
traps~\cite{fajans} and even magnetized, millimeter sized
disks rotating at a liquid-air interface~\cite{whitesides}. 
Even from a theoretical viewpoint, this remains a highly active
research front recently extending towards the consideration
e.g. of relative equilibria of N+1 vortices~\cite{anna} and
of vortex swarms~\cite{theo}.

On the other hand, over the last decade the consideration of
vortex and multi-vortex states has had a novel focal point of
attention and extensive applications, namely that of atomic Bose-Einstein 
condensates (BECs)~\cite{fetter1,fetter2,book1,book2,book3}. 
In the latter setting, most of the consideration has been
focused on the study of individual vortices and vortex
lattices~\cite{fetter1,fetter2,book3,chamoun}. However, clusters of
few vortices have been of interest both theoretically~\cite{castin}
and experimentally~\cite{chevy} since the early days of BEC
vortex experiments. Moreover, recent work on vortex dipoles
created either via  a quenching process through the
phase transition~\cite{dsh1,dsh2,dsh3} or through a superfluid
flow past a cylinder experiment~\cite{bpa10} and even studies of
tripoles~\cite{bagn,dsh3} or higher core vortex clusters 
have sparked a considerable effort to understand the properties
of such states. Notably, a fundamental twist, which is present
in this setting in comparison to the earlier fluid ones, is
the effect of an external, typically parabolic~\cite{book1,book2,book3},
trap inducing a precession of each of the vortices (with the direction
depending on the sign of its charge). It is the delicate interplay
of this precession with the vortex-vortex interaction which 
constitutes the source for numerous unexpected features in this
system, such as the presence of neutral equilibria~\cite{dsh2} for
opposite charge vortices, or
the symmetry-breaking bifurcation destabilizing symmetric
same charge vortex states.

Our emphasis in the present work will be on the study of
the so-called vortex tripole, a three-vortex configuration
in which two of the vortices are of one charge, while the
third is of the opposite charge. This configuration has been
observed experimentally in the dynamics of~\cite{bagn} (see, in 
particular, Figs. 1b, 2b and 3 therein), which are the motivating
starting point for the present considerations. Moreover, contrary
to the two-vortex system, which has been argued in~\cite{dsh2}
to be integrable (at the particle level and near-integrable
at the mean-field partial differential equation -PDE- level) 
in isotropic BECs, the three-vortex system is {\it generically}
non-integrable. This is what offers, in turn, the theoretical motivation
for the study. This setting provides one of the most elementary 
contexts, where the transition of the vortex dynamics from 
ordered states and stable periodic orbits to unstable
ones and eventually to chaotic dynamics takes place. Our scope is to provide
a systematic view toward this transition, using the integrals of
the (vortex) motion as our control parameters. It turns out that when the motion is close to the one- or two-vortex regime the motion is usually regular, while if all of the vortices interact strongly with each other the motion is general chaotic.

Our presentation is structured as follows. In section II, 
we provide the details of the mathematical model used for the study of the system of interacting vortices and its Hamiltonian formulation. After that we provide the transformations
needed in order to bring it to a reduced form, with its angular momentum $L$ as a parameter. In section III,
we proceed to the numerical study of our model by computing Poincar{\'e} 
sections 
and indicating particular bifurcations as our control parameter $L$ is varied.
In section IV, we will present a comparison of the ODE results of
our reduced equations with the corresponding PDE results of the 
Gross-Pitaevskii equation (GPE) in order to examine the
relevance of our findings based on the vortex particle model for the
physical system of interest.
Finally, in section V, we summarize our findings and present
our conclusions and some future challenges.

\section{Mathematical model of the system of interacting vortices - Hamiltonian formulation}

As it has been illustrated in the earlier works of~\cite{dsh2,dsh3}, the reduction 
of the vortex dynamics from the original experiment to that of the
mean-field PDE and from there to the ``particle'' ordinary differential
equations (ODEs) works well in suitable quasi-two-dimensional
(pancake-shaped) BECs with sufficiently large atom numbers. 
It is for that reason that, 
in what follows, we will restrict our considerations
to the case of the ODE description of the motion of the vortex cores.

A single vortex
in a harmonic trap is well-known to precess around the center
of the trap~\cite{fetter1,fetter2}. 
The frequency $\omega _{\mathrm{pr}}$ of the associated precession 
has been shown through suitable asymptotic considerations~\cite{fetter1}
(and more recently also through rigorous analysis~\cite{smets}) to depend on the parameters of the system, such as the ratio of the trap frequencies 
in the radial and z-direction
$\Omega=\omega_r/\omega_z$ and
the chemical potential $\mu$, which is directly associated with the atom number
of the 2d isotropic BEC, as well as on the distance of the
vortex from the center of the trap, $r$.  An expression that has been
argued~\cite{dsh1} to yield a good match between theory and experiment
(at least for vortices not very close to the ``outer rim'' of the 
condensate~\cite{dsh3}) is of the form:
\begin{equation}
\omega _{\mathrm{pr}}=\frac{\omega _{pr}^{0}}{1-\frac{r^{2}}{R_{\mathrm{TF}%
}^{2}}},  \label{prec}
\end{equation}%
where $R_{%
\mathrm{TF}}=\sqrt{2\mu }/\Omega $ is the so-called 
Thomas-Fermi radius, approximately characterizing the radial extent of the BEC;
$\omega _{\mathrm{pr}}^{0}$ is the precession frequency at the trap
center for which the expression
$\omega _{\mathrm{pr}}^{0}=\ln \left( A\frac{\mu }{\Omega }\right) /R_{%
\mathrm{TF}}^{2}$, with $A=2 \sqrt{2} \pi$, 
has been argued to yield good agreement with 
both PDE direct simulations and linearization spectral analysis
via the Bogolyubov-de Gennes equations~\cite{middel10}. 

On the other hand, in the absence of a harmonic trap, two interacting
vortices will rotate around each other with a frequency of $\omega _{\mathrm{%
vort}}=B/r _{ij}^{2}$, where $r_{ij}=\sqrt{\left(
x_{i}-x_{j}\right) ^{2}+\left( y_{i}-y_{j}\right) ^{2}}$ is the distance
between the vortices and $B$ is a constant factor. 
In the realm of a homogeneous BEC, $B=2$, while in the presence
of the trap, a factor lower than the value of the homogeneous case, has been used~\cite{middel10} to emulate the more complex effect of the
modulated density induced screening. A more detailed 
functional form 
(but bearing an integral kernel expression for the interaction) has been given
in~\cite{mcendoo}. However, for the considerations herein, we will
restrict ourselves to a constant $B$, following the earlier 
works of~\cite{dsh2,dsh3}, which accurately captured experimentally counter-rotating
and co-rotating vortex dynamics, respectively, through such an approach.
 This approximation is valid for
vortices that are sufficiently well-separated and thin-core (i.e., ``particle-like'') so that their structure does not affect their
inter-particle interaction. 

Based on the above assumptions, 
let us now consider $N$ interacting vortices. If $\left( x_{i},y_{i}\right)$ is the position of the $i$-th vortex, the corresponding equations of motion 
due to the other vortices and the harmonic trap are then given by~\cite{middel10,dsh2,dsh3}\
\begin{eqnarray}
\dot{x}_{i} &=&-S_{i}\omega _{\mathrm{pr}}y_{i}-B\sum_{j=1,j\neq i}^{N}S_{j}%
\frac{y_{i}-y_{j}}{2r _{ij}^{2}},  \label{middlecamp_x} \\
\dot{y}_{i} &=&S_{i}\omega _{\mathrm{pr}}x_{i}+B\sum_{j=1,j\neq i}^{N}S_{j}%
\frac{x_{i}-x_{j}}{2r _{ij}^{2}},  \label{middlecamp_y}
\end{eqnarray}%
where $S_{i}$ is the charge of the $i$-th vortex and $N$ is the total number of
interacting vortices.

We can further rescale time to the
period of the single vortex precessing near the center of the trap and space
is scaled to the Thomas-Fermi Radius according to:
\begin{equation}
\quad t\mapsto\frac{t}{\omega _{\mathrm{pr}}^{0}},\quad x\mapsto xR_{\mathrm{TF}},\quad y\mapsto{y}{R_{\mathrm{TF}}}.  \label{dimtxy}
\end{equation}

Then, the resulting equations of motion read:
\beq\begin{array}{rrl}
\dot{x}_i=&\ds-S_i\frac{y_i}{1-r_i^2}&\ds-c\sum_{j=1, j\neq i}^NS_j\frac{y_i-y_j}{r_{ij}^2}\\[12pt]
\dot{y}_i=&\ds S_i\frac{x_i}{1-r_i^2}&\ds+c\sum_{j=1, j\neq i}^NS_j\frac{x_i-x_j}{r_{ij}^2},
\end{array}\label{eq_mot_norm}\eeq
where we have introduced the non-dimensional parameter
\begin{equation}
c=\frac{B}{2\ln \left( A\frac{\mu }{\Omega }\right) }.
\end{equation}

For our unit charge vortices, this dynamical evolution can be acquired by the Hamiltonian
\beq H=\frac{1}{2}\sum_{k=1}^N\ln(1-r_k^2)-\frac{c}{2}\sum_{k=1}^N\sum_{j>k}^NS_kS_j\ln(r_{kj}^2)\label{hamiltonian_general}\eeq
through the canonical equations
$$\dot{x}_i=S_i\frac{\pa H}{\pa y_i}\, ,\quad \dot{y}_i=-S_i\frac{\pa H}{\pa x_i}\ .$$
In what follows hereafter, we restrict our 
study to the case of interest, namely the tripole with two vortices of one circulation and one of opposite circulation.
 Vortices are assumed to be of unit charge, since these are generically stable, contrary to the unstable
case for higher charges~\cite{pu}.

\subsection{The $N=3$, $S_1=S_3=1$, $S_2=-1$ case}
We will consider a system of $N=3$ interacting vortices, two 
with charge $S_1=S_3=1$ and one with charge $S_2=-1$, as per
the observations of~\cite{bagn}. According to (\ref{hamiltonian_general}), the Hamiltonian in this case will be
$$H=\frac{1}{2}\sum_{i=1}^3\ln(1-r_i^2)+\frac{c}{2}\left[\ln(r_{12}^2)-\ln(r_{13}^2)+\ln(r_{23}^2)\right],$$
where we recall that $r_i=\sqrt{x_i^2+y_i^2}$ and $r_{ij}=\sqrt{\left(
x_{i}-x_{j}\right) ^{2}+\left( y_{i}-y_{j}\right) ^{2}}$. If we define $\mathbf{q}=(x_1, y_2, x_3)$ to be the {\it generalized positions} of the system and  $\mathbf{p}=(y_1, x_2, y_3)$ to be the conjugate {\it generalized momenta}, the corresponding equations of motion (\ref{eq_mot_norm}) are derived through the 
standard 
Hamilton's canonical equations

$$\dot{q}_i=\frac{\pa H}{\pa p_i}\ ,\quad \dot{p}_i=-\frac{\pa H}{\pa q_i}.$$ 
\noindent {\bf Remark:} Let ${\bm\eta}=(\mathbf{q},\ \mathbf{p})^T=(x_1,\ y_2,\ x_3,\ y_1,\ x_2,\ y_3)^T$. Then the equations of motion can be written as $$\dot{{\bm\eta}}={\bm \Omega} D_{\bm{\eta}}H,$$
where $\Omega$ is the standard matrix of the symplectic structure $\Omega=\left(\begin{array}{cc} \mathbf{O}&\mathbf{I}\\ -\mathbf{I}& \mathbf{O}\end{array}\right)$, with $\bf I$ and $\bf O$ being the 
$3\times 3$ identity and zero matrices respectively. By $D_{\bm{\eta}}$ we denote the $({\pa}/{\pa_{\eta_1}}\ldots{\pa}/\pa_{\eta_6})^T$ operator.
Note 
that, one could use a different arrangement of variables, say ${\bm \eta}'$. The system 
would still have a Hamiltonian structure but with a 
different symplectic matrix ${\bm \Omega}'$. 
We choose the above mentioned arrangement of variables because it is convenient for the computation of the Poincar\'e sections.

\subsection{The reduced Hamiltonian} 

The Hamiltonian of the system can be reduced to a two-degree of freedom one, 
by applying two canonical transformations (i.e. transformations which preserve the functional form of the equations of motion). 
The first transformation concerns the rewriting of the Hamiltonian using the Poincar\'e variables $(w_i,R_i)$, which are defined by
\beq q_i=\sqrt{2R_i}\sin(w_i)\ ,\quad p_i=\sqrt{2R_i}\cos (w_i)\label{can_transf_1}.\eeq
Then, the Hamiltonian of the system becomes
$$\begin{array}{rcl}
H&=&\ds\frac{1}{2}\left[ \ln(1-2R_1)+\ln(1-2R_2)+\ln(1-2R_3) \right]+\\[10pt]
 &+&\ds\frac{c}{2}\left[ \ln(2R_1+2R_2-4\sqrt{R_1R_2}\sin(w_1+w_2))-\ln(2R_1+2R_3-4\sqrt{R_1R_3}\cos(w_1-w_3))+\right.\\[10pt]
& & \ds\quad\left. +\ln(2R_2+2R_3-4\sqrt{R_2R_3}\sin(w_2+w_3)), \right]
\end{array}$$

where the $R_i$'s must satisfy $R_i<0.5$, which is tantamount to the vortices being located within the Thomas-Fermi radius. From the form of the transformed Hamiltonian it is obvious that the proper variables to use are not $w_i$ but linear combinations thereof. So, we apply a second canonical transformation 
\beq\begin{array}{rlcrl}
\phi_1=&w_1-w_3&\quad\quad&J_1=&R_1\\
\phi_2=&w_2+w_3&\quad\quad&J_2=&R_2\\
\thet=&w_3&\quad\quad&L=&R_1-R_2+R_3.
\end{array}\label{can_transf_2}\eeq

In this new set of variables, $J_1,\ J_2,\ L$ are the conjugates of $\phi_1,\ \phi_2, \thet$ respectively and have the specific chosen form in order for the transformation to be canonical. The variables $\phi_1,\ \phi_2$ can be considered as {\it generalized phase differences} (which differ from the standard ones because of the sign of $S_2=-1$) while $\thet$ can be considered as the {\it phase} of the system. Since $\thet$ is an angle variable, its conjugate variable $L$  is the 
corresponding {\it angular momentum}. In general, the
angular momentum for a vortex system is defined as $L= \sum_i S_i r_i^2$~\cite{aref1,newton1}, which, by using (\ref{can_transf_1}), coincides with the one in (\ref{can_transf_2})
and can be shown via direct computation to be a conserved quantity
for the dynamical system associated with 
Eqs.~(\ref{middlecamp_x})-(\ref{middlecamp_y}).

By using the above mentioned transformation the Hamiltonian becomes

\beq\begin{array}{rcl}
H&=&\ds\frac{1}{2}\left[ \ln(1-2J_1)+\ln(1-2J_2)+\ln(1-2(L-J_1+J_2)) \right]\\[10pt] 
 &+&\ds\frac{c}{2}\left[ \ln(4J_2-2J_1+2L-4\sqrt{J_2}\sqrt{L-J_1+J_2}\sin(\phi_2))-\ln(2L+2J_2-4\sqrt{J_1}\sqrt{L-J_1+J_2}\cos(\phi_1))\right.\\[10pt]
& & \ds\quad\left. +\ln(2J_1+2J_2-2\sqrt{J_1}\sqrt{J_2}\sin(\phi_1+\phi_2)) \right].\\[10pt]
\end{array}\label{hfJ}\eeq

Since the variable $\thet$ is ignorable (it is not explicitly contained in the Hamiltonian), the corresponding variable $L$ is, as expected, an 
integral of motion. So, the system is transformed into a 2 degrees-of-freedom 
Hamiltonian system with the angular momentum $L$ as a parameter. Since it appears that no other integral of motion exists (in addition to the $H$ and $L$), the Hamiltonian (\ref{hfJ})
is non-integrable. It is evident that the system represents the simplest non-integrable
dynamical variant within an isotropic two-dimensional BEC. Thus, it is expected that regular and chaotic motion will coexist in the phase space of the system. So, the natural consideration is to study the transition of the dynamics from completely regular to progressively chaotic\footnote{Note here that a Hamiltonian 
system can never exhibit completely chaotic behavior since there are always islands of regularity in its phase space.}, as parameters, such as the angular momentum $L$, are varied. For two degree-of-freedom systems, the Poincar\'e sections are the most illustrative tool for the investigation of the underlying dynamics. 

\section{Phase space exploration through Poincar\'e sections.} 

We will study the dynamical behavior of this tripole system by using a sequence of Poincar\'e sections. 
A Poincar\'e section is defined for a fixed value of the energy of the system, which could be determined by some particular initial conditions of the motion, i.e. $h=H(x_{10},\ y_{10},\ x_{20},\ y_{20},\ x_{30},\ y_{30})$. So, in order to span different relative position settings for the
vortices of energy $h$, we will consider various values of the angular momentum $L$, which is used as a 
constant parameter for each section. In our numerical computations, the value of the energy has been chosen to be $h=-0.7475$, which corresponds to a typical initial configuration of the vortices as it it is shown in Fig.~\ref{fig:x_y_po_L_m_0_25}. On the other hand, as it can be shown from (\ref{can_transf_2}) and by the fact that $R_i$ lie in the range $0<R_i<0.5$ we conclude that $L$ can vary in the range $-0.5<L<1$. But, since there is the energy constraint as well, this range will be actually smaller.

For the surface of section we consider the ($\phi_1-J_1$) plane and fix the value of $\phi_2$ to be $\phi_2=\pi/2$. This value of $\phi_2$ corresponds to the state where the $S_2$ and $S_3$ vortices lie on the half-line having the center of the condensate on its edge as it can be seen from the transformations (\ref{can_transf_1}) and (\ref{can_transf_2}). Then, the value of $J_2$, for a given pair of $\{\phi_1, J_1\}$ on the plane, 
can be inferred by solving the equation $h=H(\phi_1,\ J_1,\ \phi_2,\ J_2,\ L)$. 

It is reminded here that, in general, isolated (fixed or periodic) points on the 
surface of section correspond to periodic orbits of the particular dynamical system. Similarly,
invariant curves in the Poincar\'e sections will correspond to quasi-periodic
orbits. But, this applies to the ``reduced'' phase-space of the $\{\phi_1, \phi_2, J_1, J_2\}$ variables. In the present setting we have eliminated the third degree of freedom, by using the angular momentum integral $L$.
This additional ``hidden'' degree of freedom introduces an extra frequency which is not necessarily commensurate with the frequencies of the other two degrees of freedom.
As a result, a fixed point of the Poincar\'e section does not correspond, in general,
to a periodic orbit in the ($x-y$) coordinates of the ``full'' system but to a {\it quasi-periodic}
one. As we embark on a detailed analysis of the different orbits,
this additional complication (bearing an additional frequency for each orbit
to those ``normally'' counted on the basis of the Poincar\'e section)
should be borne in mind. Moreover, the term orbit is used in order to describe a trajectory in three different spaces; the full ($x-y$) phase-space, the reduced \{$\phi_1, \phi_2, J_1, J_2$\} phase-space and the ($\phi_1-J_1$) surface of section. But, the distinction between them should be clear depending on the context.

\subsection{The regular motion region ($L<L^*$)}
\begin{figure}[htbp]
	\centering
		\includegraphics[width=12cm]{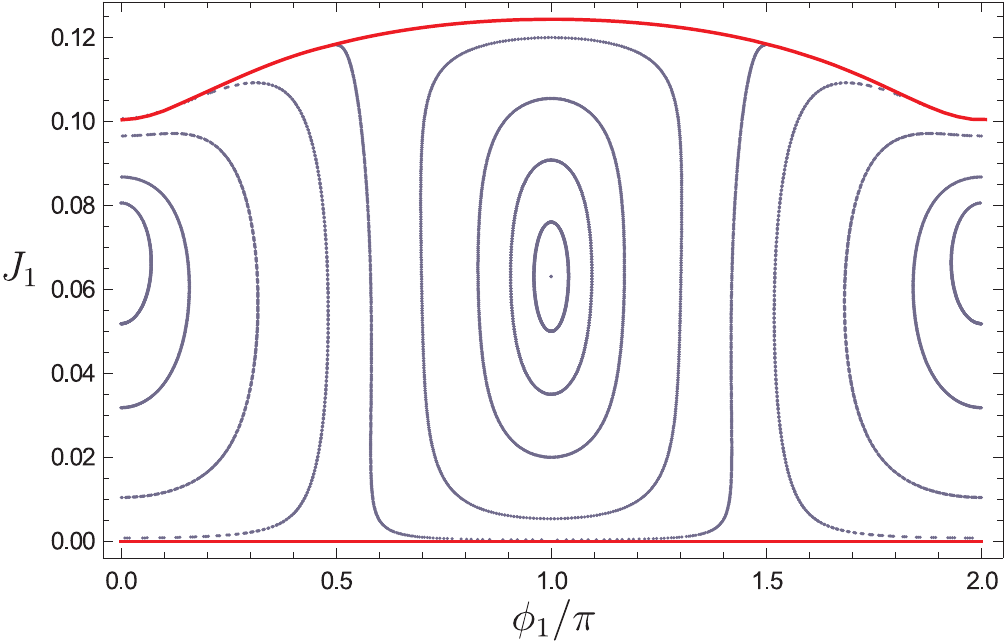}
	\caption{(Color Online) The Poincar\'e section for $L=-0.25$. We can distinguish two regions of predominantly ordered dynamics, as 
discussed in detail in the text. The red (light colored) 
lines depict the boundaries of motion.}
	\label{fig:sec_L_m_0_25}
\end{figure}

\begin{figure}[htbp]
	\centering
		\includegraphics[width=5.cm]{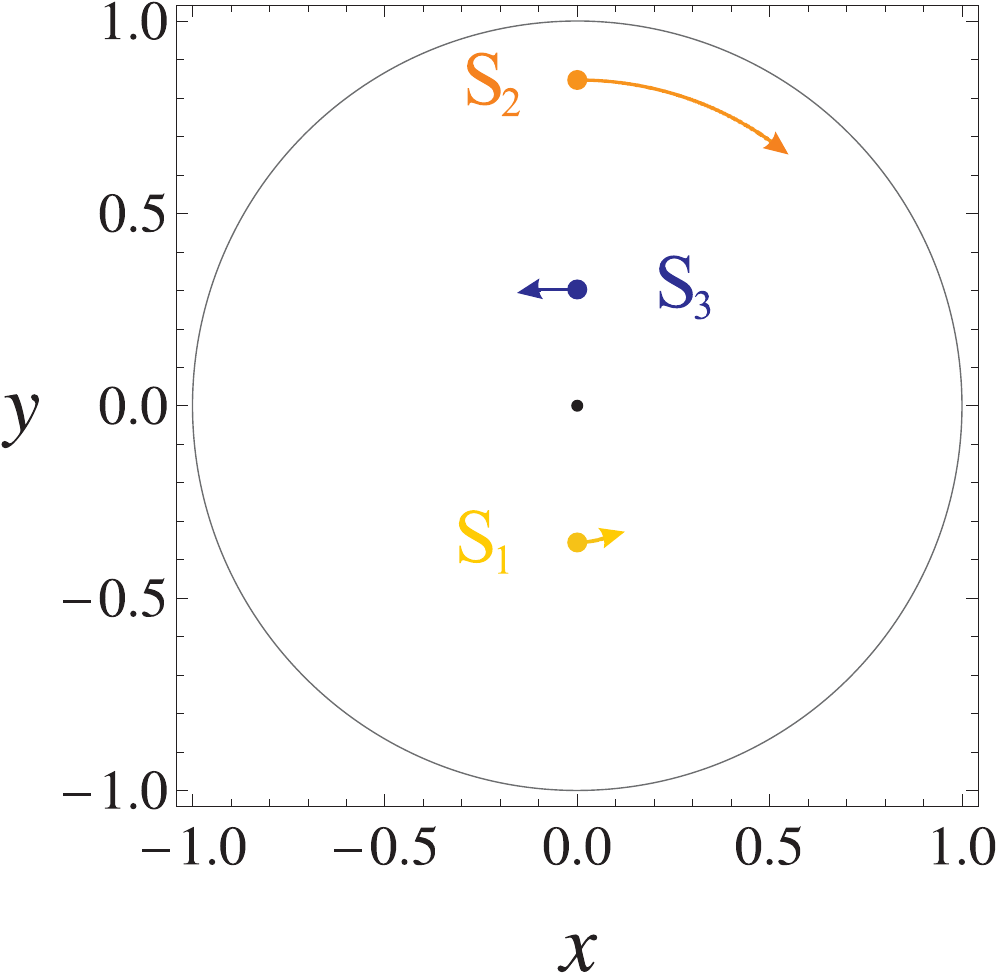}
		\includegraphics[width=5.cm]{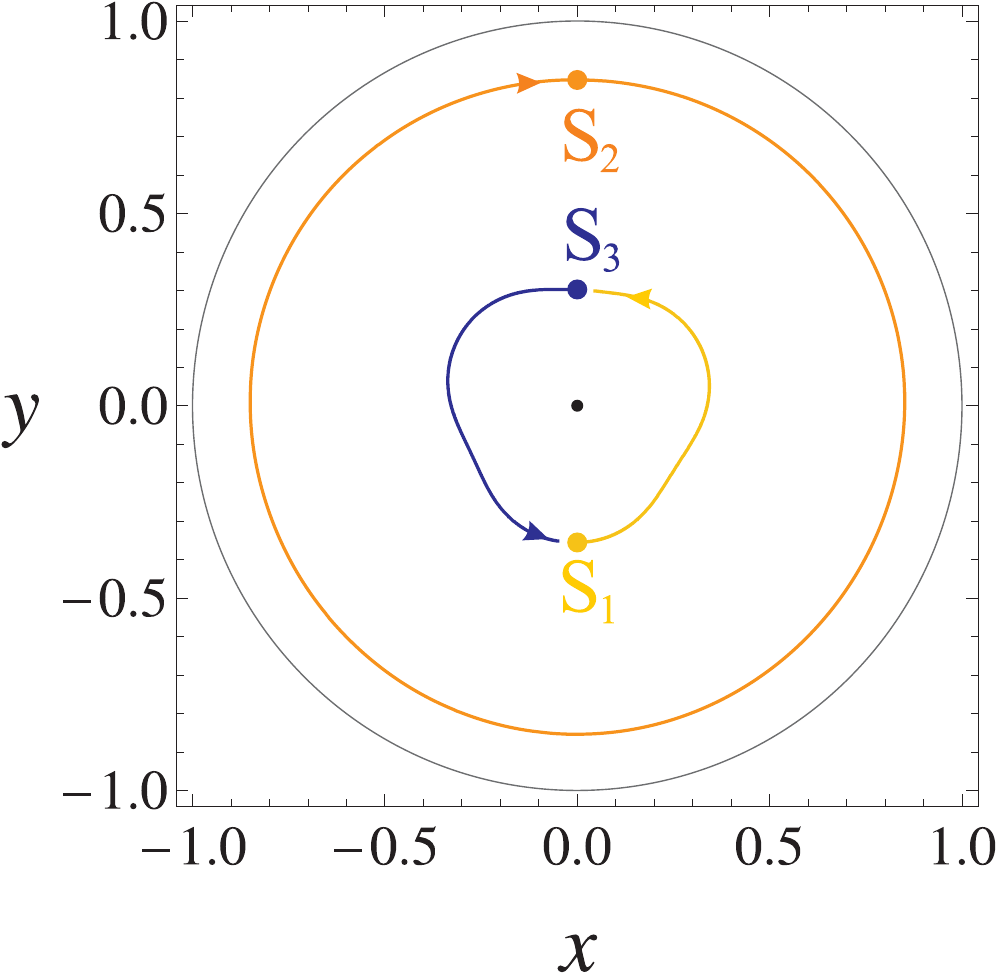}
		\includegraphics[width=5.cm]{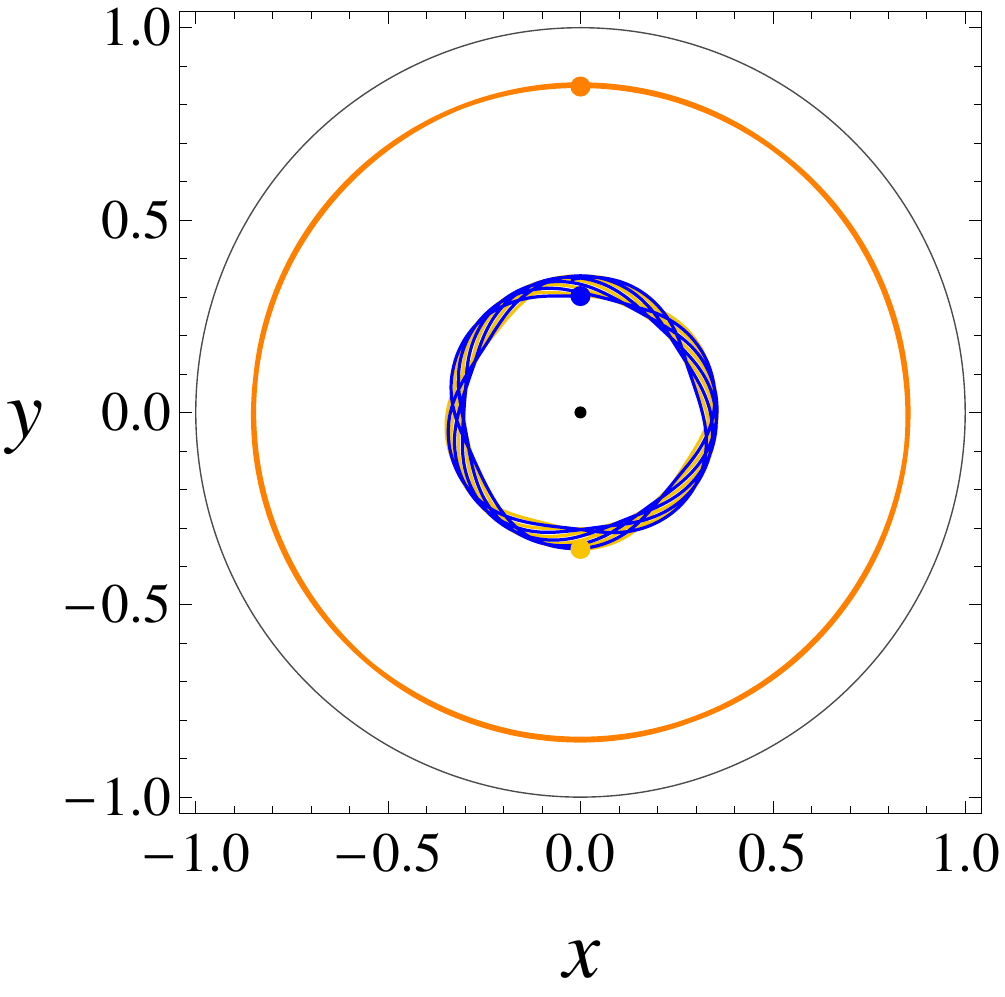}
	\caption{(Color Online) An orbit in the ($x-y$) space for $L=-0.25$ is shown. It corresponds to the ``central'' periodic orbit of Fig.\ref{fig:sec_L_m_0_25}, which in turn represents a quasi-periodic orbit in the $(x-y)$-space. Here it is shown for three values of the time, $t=0.2$, $t=1.9$, $t=20$, from left to right. The dots are showing the initial configuration. The yellow, orange and blue (light, intermediate and dark grey) dots correspond to the $S_1$, $S_2$ and $S_3$ vortices respectively. The same same colors are used to the solid lines which denote the corresponding trajectories. The black dot represents the center of the condensate, while the gray circle represents the Thomas-Fermi radius.. This color code will be followed for the rest of this work. In addition, in the first two panels there are arrows indicating the main rotation direction of the vortices which is dictated by the gyroscopic precession due to the trapping potential.}
	\label{fig:x_y_po_L_m_0_25}
\end{figure}

For all $L<-0.218=L^*$ the basic dynamical behavior is essentially
invariant. In Fig.~\ref{fig:sec_L_m_0_25} we show the Poincar\'e section for $L=-0.25$ as a representative of this regime of angular momenta. 
In this figure we can distinguish two regions. 
The first region consists of regular (quasi-periodic) orbits
around the central periodic orbit with 
coordinates $(\phi_1, J_1)\simeq(\pi, 0.063)$. This region 
corresponds to the ``rotational'' regime where all three vortices rotate around the center 
of the condensate, each one having the rotation direction which is determined 
by its corresponding charge (counter-clockwise for the positive charges, 
clockwise for the negative charge). As regards the periodic orbit
at $(\phi_1, J_1)\simeq(\pi, 0.063)$, its true quasi-periodic nature in the ($x-y$) plane 
is revealed in Fig.~\ref{fig:x_y_po_L_m_0_25}, where it is evident that it does not repeat itself, but instead covers densely a specific area of the 
plane. Furthermore, such orbits where $\phi_2=\pi/2$ and 
$\phi_1=\pi$ correspond to configurations having all the vortices aligned with the rotation center and having the  $S_2$, $S_3$ vortices on the opposite side of $S_1$. This kind of configurations will be called {\it symmetric} in what follows.

\begin{figure}[htbp]
	\centering
         	\includegraphics[width=5.cm]{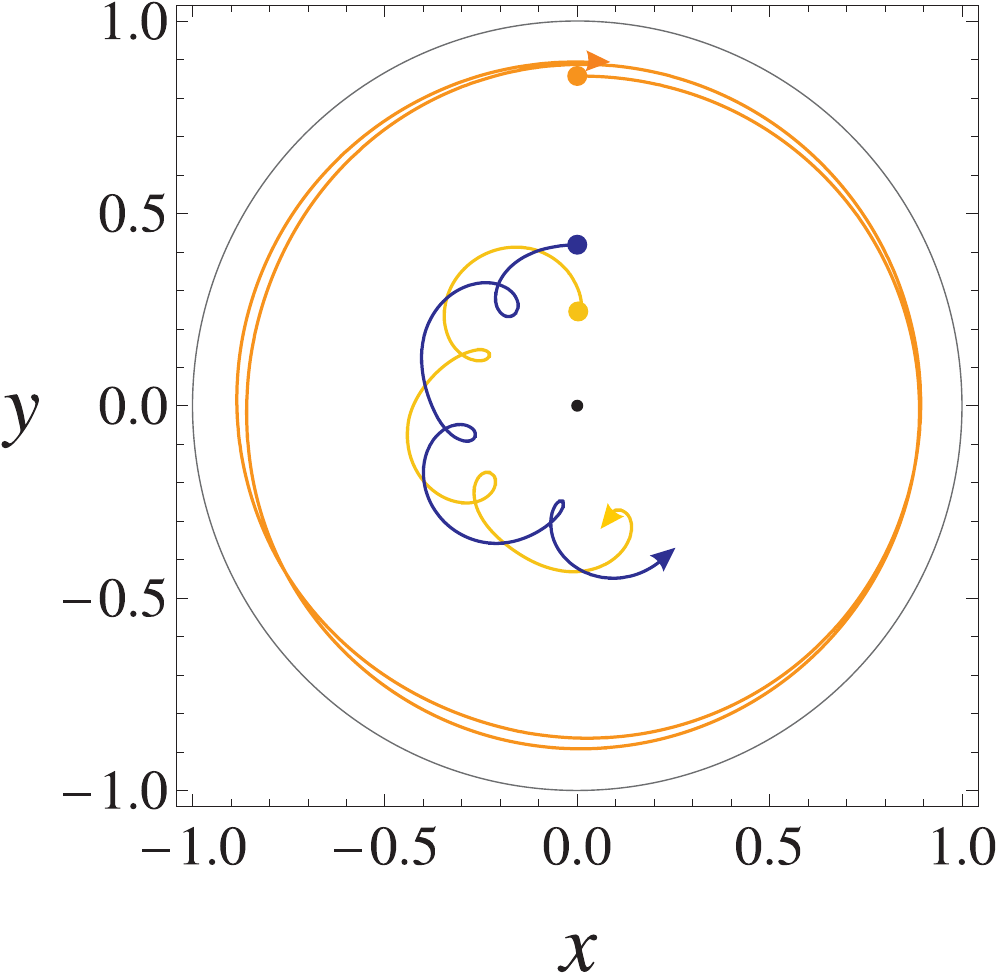}\includegraphics[width=5.cm]{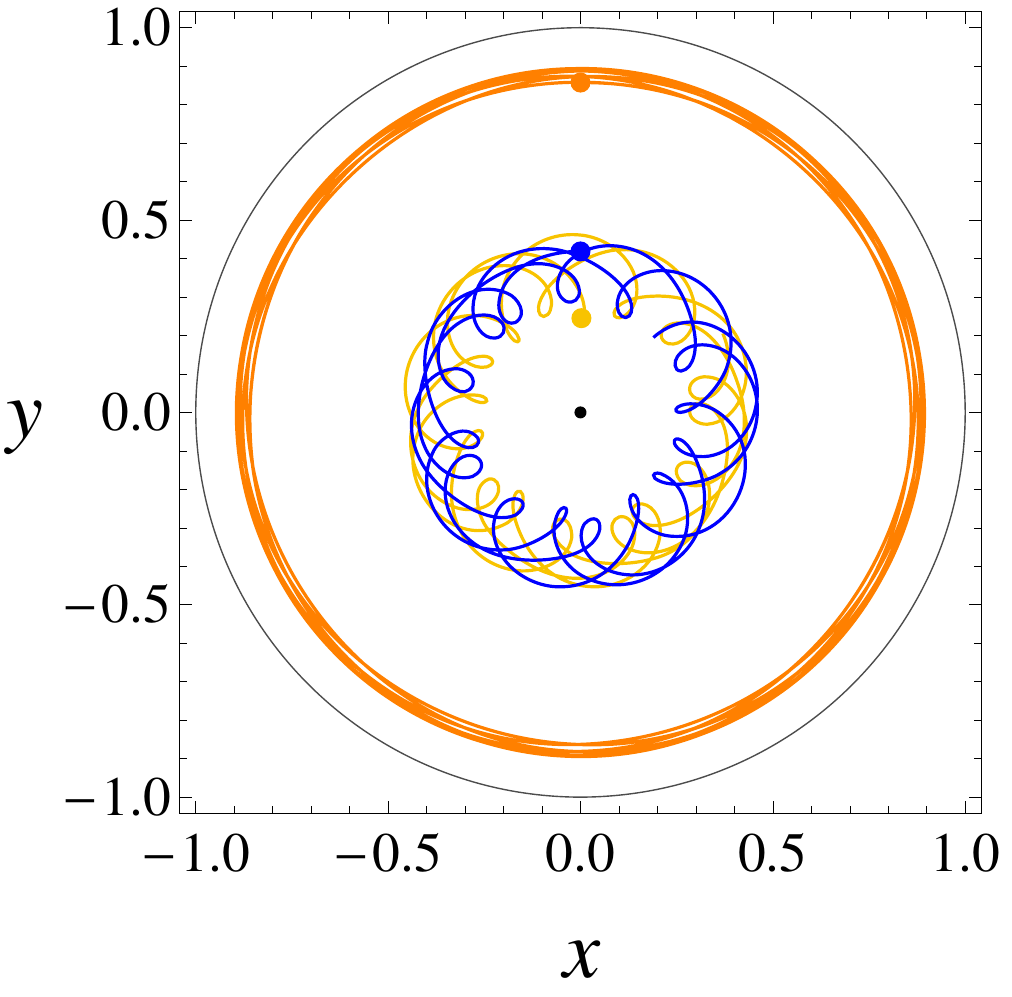}
	\caption{(Color Online) An orbit for $L=-0.25$ is shown which is a representetative of the ``satellite'' regime orbits. It corresponds to $(\phi_1,\ J_1)\simeq(0,\ 0.03)$ and it is depicted for $t=3$ (left panel) and $t=10$ (right panel).}
	\label{fig:x_y_nc_L_m_0_25}
\end{figure}

The second distinct region of motion in the Poincar\'e section is the set of regular orbits around $(\phi_1,\ J_1)\simeq(0,\ 0.07)$. Note that, since $\phi_1$ is cyclic, the topology of the section is cylindrical and not flat. 
One could have expected this point to correspond to a periodic orbit, since it is surrounded by quasi-periodic orbits, but instead a 
{\it collision} between the $S_1$ and $S_3$ vortices occurs. The motion on 
the $(x-y)$ plane of a characteristic orbit ($(\phi_1,\ J_1)\simeq(0.1,\ 0.03)$) in this area is shown in Fig.~\ref{fig:x_y_nc_L_m_0_25}.
This region corresponds to the motion where the $S_1$ and $S_3$ particles rotate around each other and both of them around the center. Such kind of motion is characterized as the ``satellite'' regime and constitute a very good representative of the two-vortex regime which is discussed in section \ref{last}.

In Fig.~\ref{fig:sec_L_m_0_25}, there are also two red curves which represent the boundaries of the Poincar\'e section and correspond to the configurations in which one of the vortices is located at the center of the rotation. The shape of these curves is calculated by using the requirement of the suitable vortex to have $R=0$. In particular, in the section of Fig.~\ref{fig:sec_L_m_0_25} we have used the conditions $R_1=0$ and $R_3=0$. This is because, as it can be seen by the transformations (\ref{can_transf_1}) and (\ref{can_transf_2}), the positive contribution to $L$ is related to $R_1$ and $R_3$, while the negative contribution comes from $R_2$. Since the section of Fig.~\ref{fig:sec_L_m_0_25} corresponds to a low value of $L$ ($L=-0.25$), it means that the $S_1$ and $S_3$ vortices are moving close to the center of rotation. More specifically, the lower limit is calculated by considering $R_1=J_1=0$ while for the upper boundary 
we consider $R_3=0$ which leads to $J2=J1-L$. By inserting the last relationship in Eq.~(\ref{hfJ}) as well as the fixed values of $h$, $L$ and $\phi_2=\pi/2$ we get the implicit formula for the function $J_1=J_1(\phi_1)$ of the upper boundary.

When the orbit of a vortex passes through the origin, the corresponding invariant curve on the Poincar\'e section collides with the boundary at a point $(\phi_1, J_1)=(\phi^*, J^*)$. At this point the angle $w$ of the corresponding vortex undergoes a discontinuous change which implies a discontinuous change of the invariant curve which continues at the point $(\phi_1, J_1)=(2\pi-\phi^*, J^*)$. Although the true motion is not interrupted, the corresponding invariant curve in the Poincar\'e section appears disconnected, which is an artifact of the particular transformation (\ref{can_transf_1}) we have used. These specific orbits, which collide with the boundary, act as ``separatrices'', distinguishing the rotational and satellite regimes.

On the other hand, the boundaries describe the upper and lower limits of the values of $J_1$ which correspond to the extreme values the distance $R_1$ can acquire satisfying also the constraint of the fixed value of $L$.
\subsection{The $L^*<L\leqslant0$ region}
\begin{figure}[h]
	\centering
		\includegraphics[width=12cm]{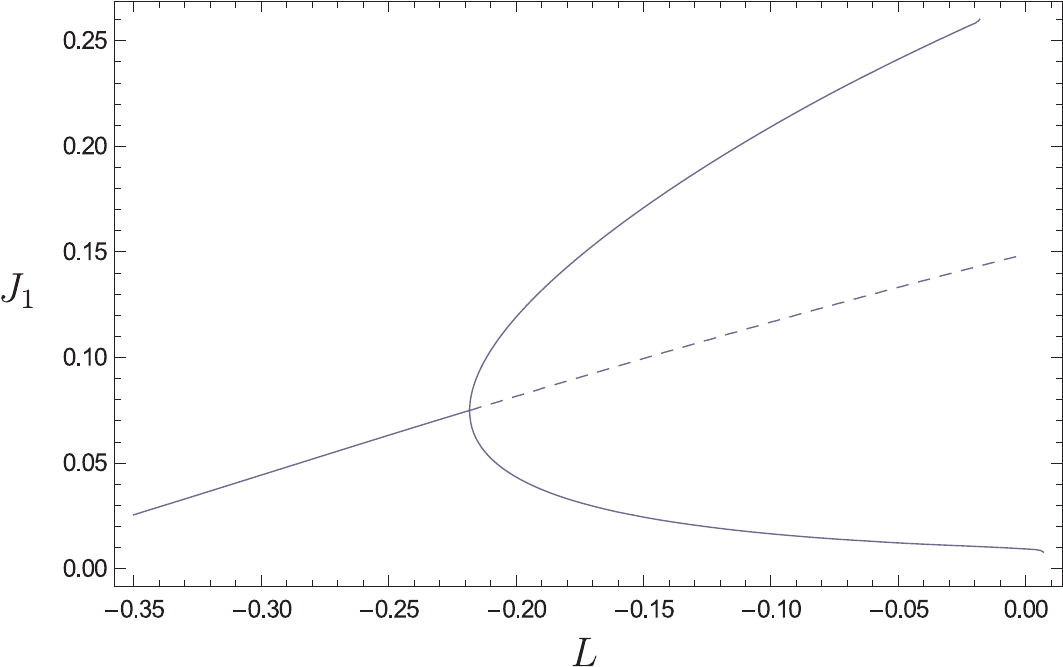}
	\centering
	\begin{tabular}{ccc}
		$L=-0.215$&\hspace{1cm}&$L=-0.2$\\[10pt]
		\includegraphics[width=7cm]{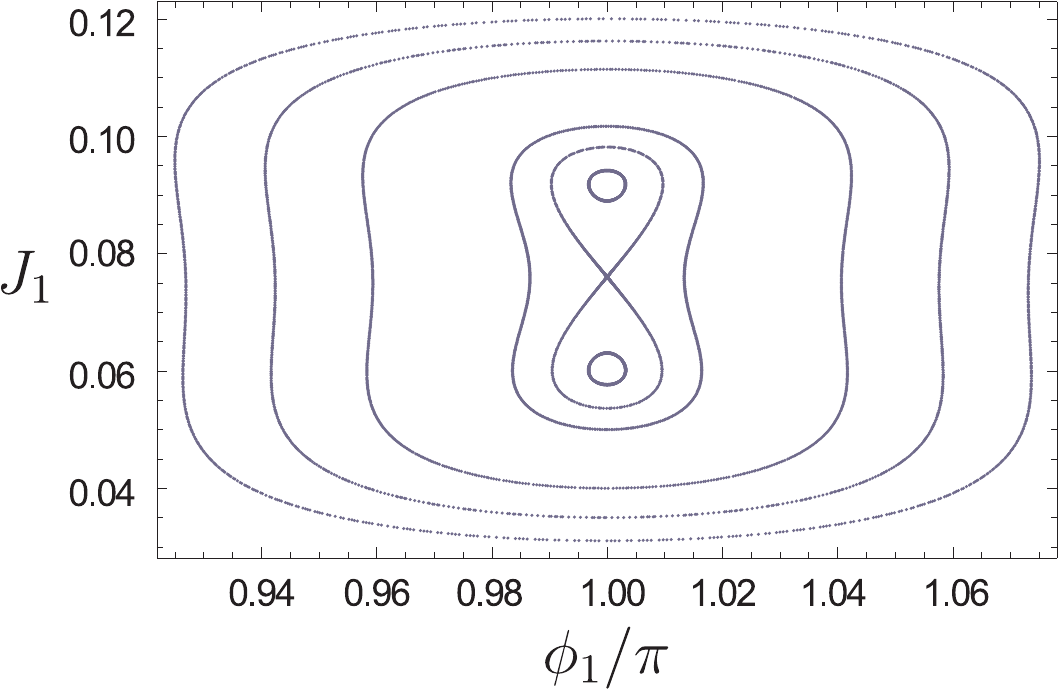}&\hspace{1cm}&\includegraphics[width=7cm]{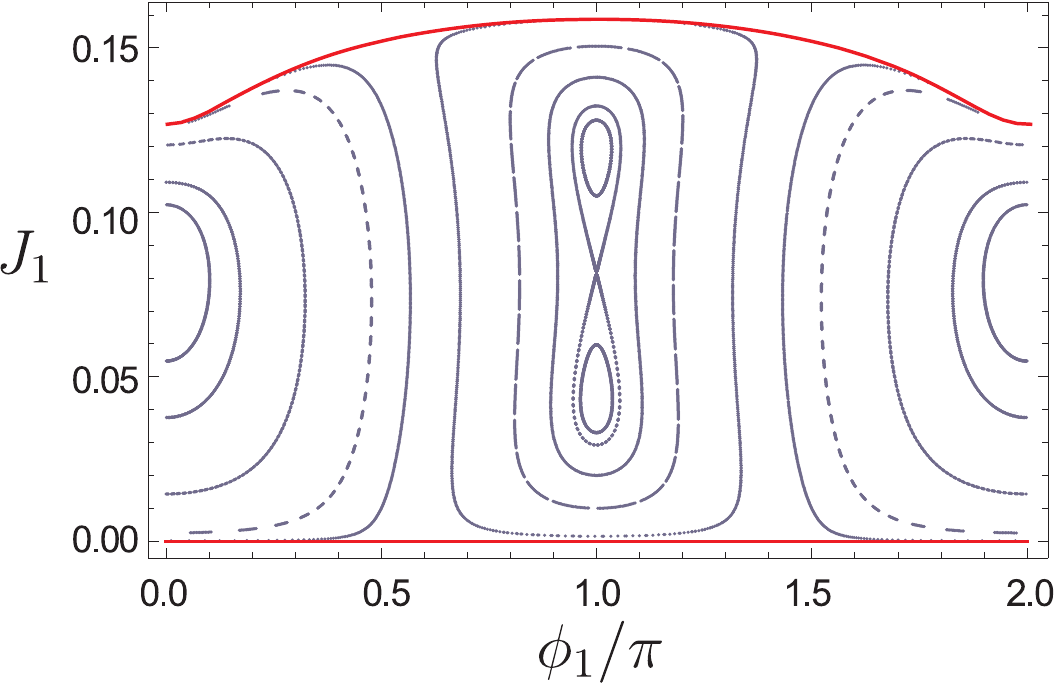}	
	\end{tabular}
	\caption{(Color online) At $L=-0.218$ a supercritical 
pitchfork bifurcation occurs, replacing the central stable periodic orbit with an unstable and two stable ones. In the diagram the value of 
$J_1$ of these orbits on the section is shown as a function of
$L$. The value of $\phi_1$ is constant and equal to $\phi_1=\pi$, 
i.e. all of them are symmetric (in the meaning of the term indicated above
for vortex orbits considered herein). 
The Poincar\'e section of the system after the first pitchfork bifurcation which occurs for $L\simeq -0.218$ is shown in the bottom panels of
the figure. The left panel corresponds to a magnification of the section 
for $L=-0.215$ around $\phi_1=\pi$. In the 
right panel we show the full section for $L=-0.2$.}
\label{fig:pitchfork}
\end{figure}

\begin{figure}[htbp]
	\centering
		\includegraphics[width=7cm]{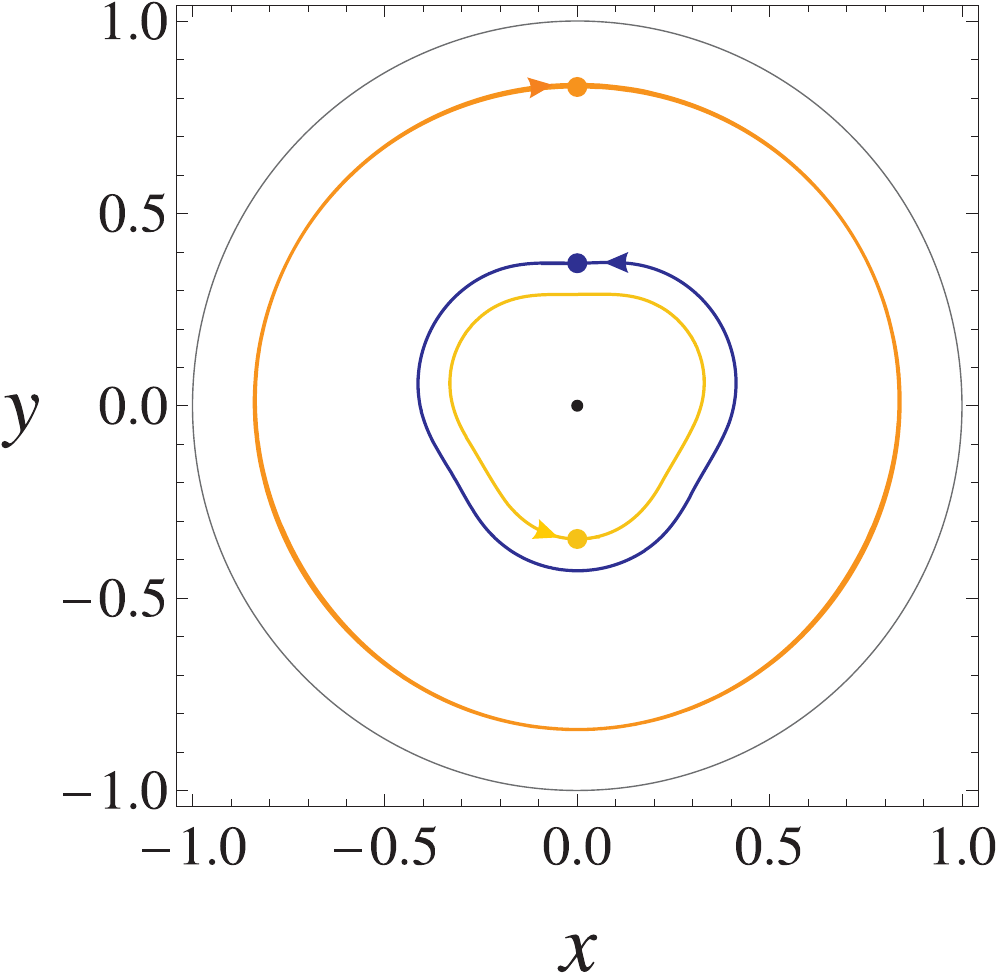}\hspace{0.7cm}\includegraphics[width=7cm]{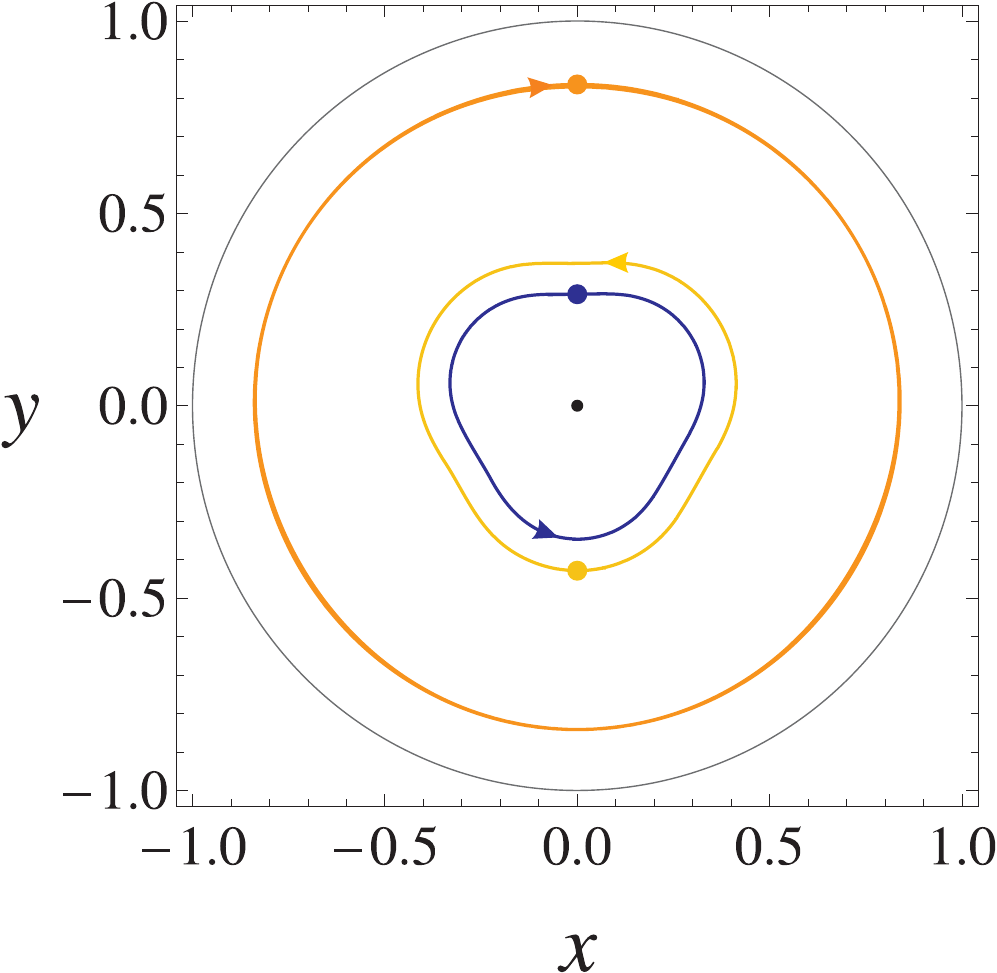}
\centering		\includegraphics[width=7cm]{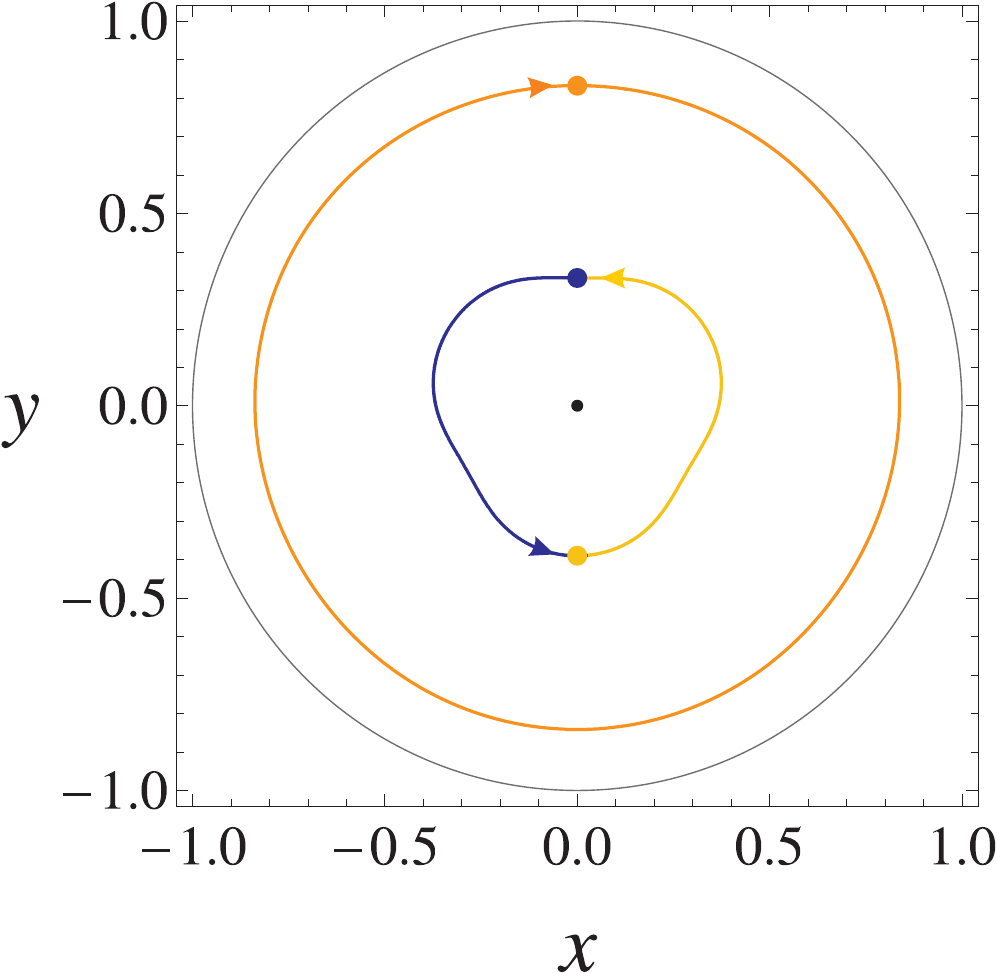}\hspace{0.7cm}\includegraphics[width=7cm]{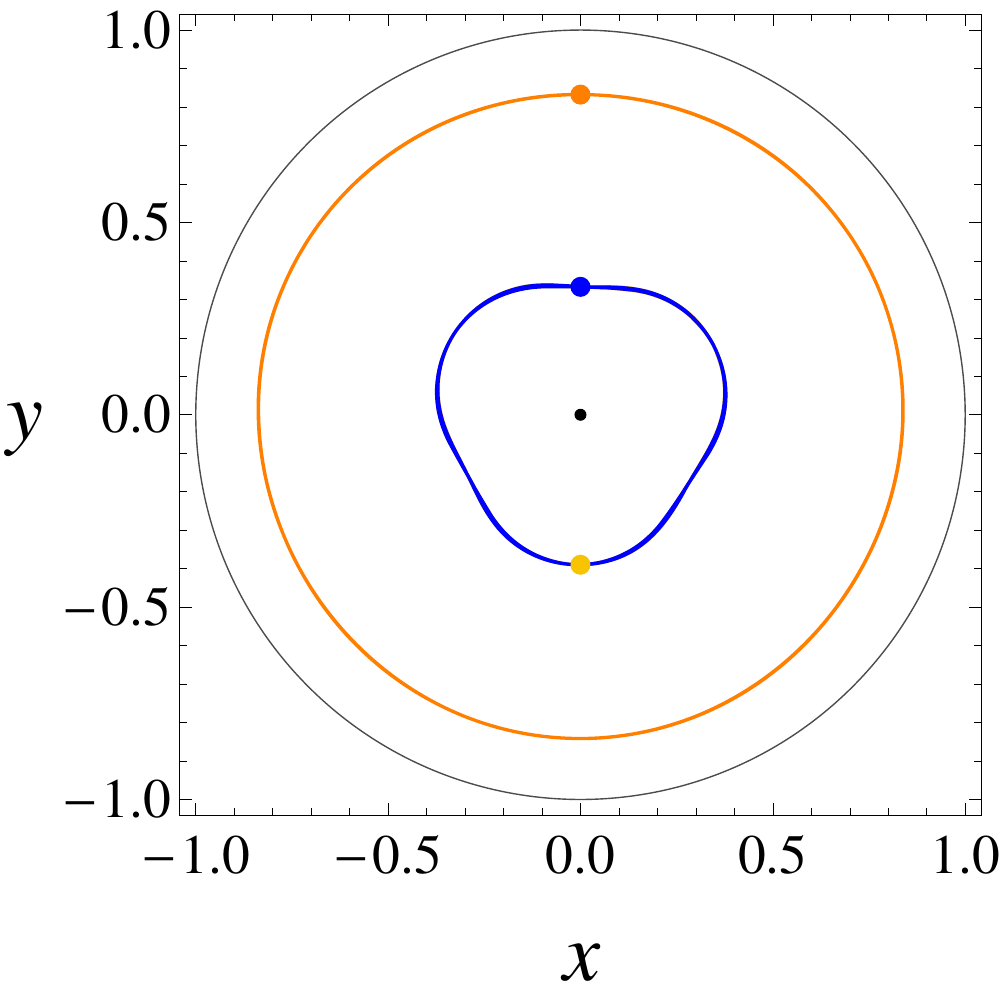}
	\caption{(Color online) Top panels: the two stable periodic orbit configurations for $L=-0.215$, for $t=2.1$. The two orbits are similar. In the one in the left 
panel which corresponds to $(\phi_1,\ J_1)\simeq(\pi,\ 0.06)$ the $S_1$ vortex 
rotates closer to the center than $S_3$, while in the configuration of the 
right panel, which corresponds to $(\phi_1,\ J_1)\simeq(\pi,\ 0.0918)$ 
the situation is reversed.
Bottom panels: the motion of the vortices which corresponds to the central 
unstable periodic orbit for $L=-0.215$. In the left panel the evolution for 
$t=2.1$ is shown while in the right hand panel the system has evolved for 
$t=20$. As we can see the orbits of the vortices lie in an almost 
one-dimensional subspace of the ($x-y$) plane, indicating 
that this orbit is very close to the ``exact'' periodic orbit.} 

	\label{fig:x_y_L_m_0_215_stable}
\end{figure}

The first structural change in the Poincar\'e sections occurs for  $L\simeq-0.218$, where the ``central'' stable periodic orbit is replaced by one unstable and two stable through a supercritical ``pitchfork'' 
(i.e., spontaneous symmetry breaking) 
bifurcation as shown in Fig.~\ref{fig:pitchfork}.
The top panel of the figure shows the location of the fixed points
of the Poincar{\'e} section, rendering evident the nature of the bifurcation;
this change in the form of the surface of section is illustrated
in the bottom panels of the figure. 
In the left panel a magnification of the section for $L=-0.215$ around $\phi_1=\pi$ is depicted in order for the bifurcation to be shown. We can clearly see that the ``central'' stable periodic orbit is replaced by an unstable 
saddle point while two stable symmetric periodic orbits appear as well. 
The two stable configurations which correspond to $(\phi_1,\ J_1)\simeq(\pi,\ 0.06)$ and $(\phi_1,\ J_1)\simeq(\pi,\ 0.0918)$ are shown in the
($x-y$) space in the top panels of Fig.~\ref{fig:x_y_L_m_0_215_stable}. As we can see, the two configurations are almost mirror images of
each other, with the left-hand one having $S_1$ orbiting more closely to the center than $S_3$, while in the right-hand panel the situation is reversed. 
In the bottom panels of Fig.~\ref{fig:x_y_L_m_0_215_stable} the unstable (symmetric) periodic orbit for
$(\phi_1,\ J_1)\simeq(\pi,\ 0.076)$ is depicted. In particular, its time evolution for $t=2.1$ and $t=20$ is shown. In these plots we can see that the area that the orbits of the vortices in the ($x-y$) plane are occupying is very narrow, almost one-dimensional. This indicates that the unstable orbit is very close to an ``exact'' periodic orbit of the full system, which actually occurs for $L\simeq -0.213$. 
In the right panel of Fig.~\ref{fig:pitchfork}, 
the full section for $L=-0.2$ is depicted. 
This also serves to illustrate that
although the picture in the central region has changed, 
the region around the edges remains
qualitatively similar as before the bifurcation.

\begin{figure}[htbp]

\begin{tabular}{ccc}	
\hspace{-1cm}	$L=-0.09$&$L=-0.07$&$L=-0.05$\\
\hspace{-1cm}\includegraphics[width=6cm]{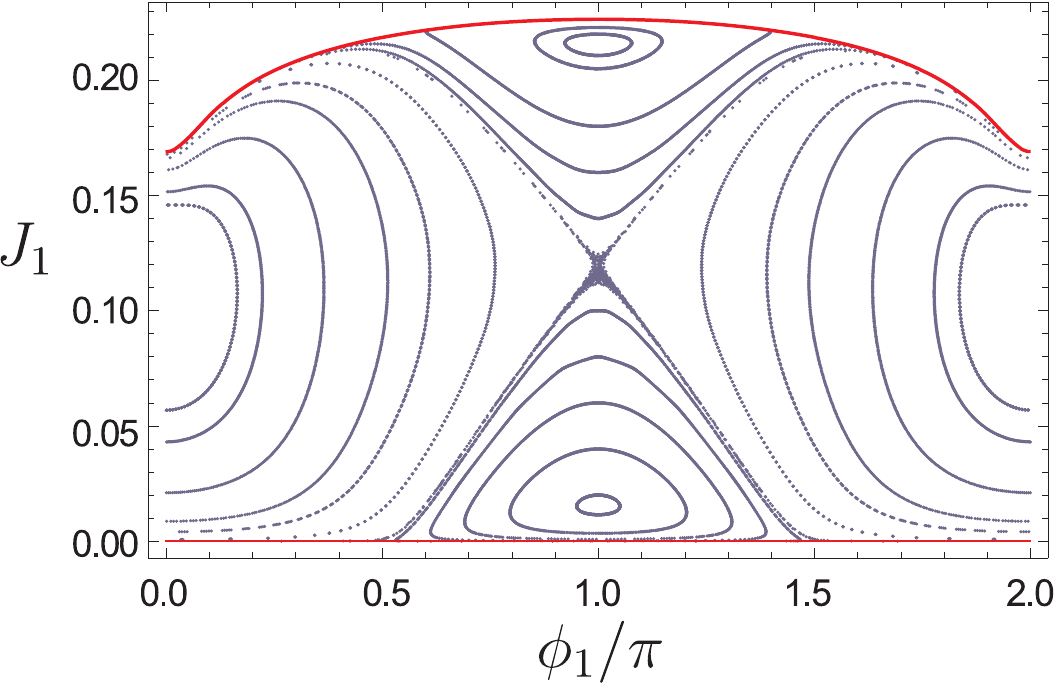}&\includegraphics[width=6cm]{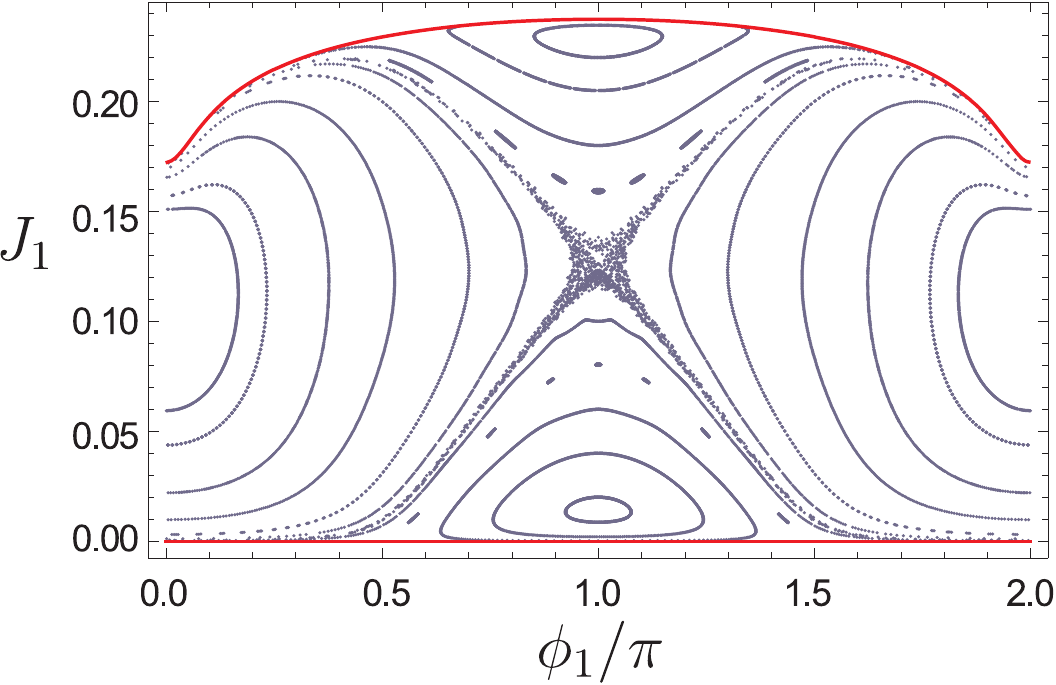}&\includegraphics[width=6cm]{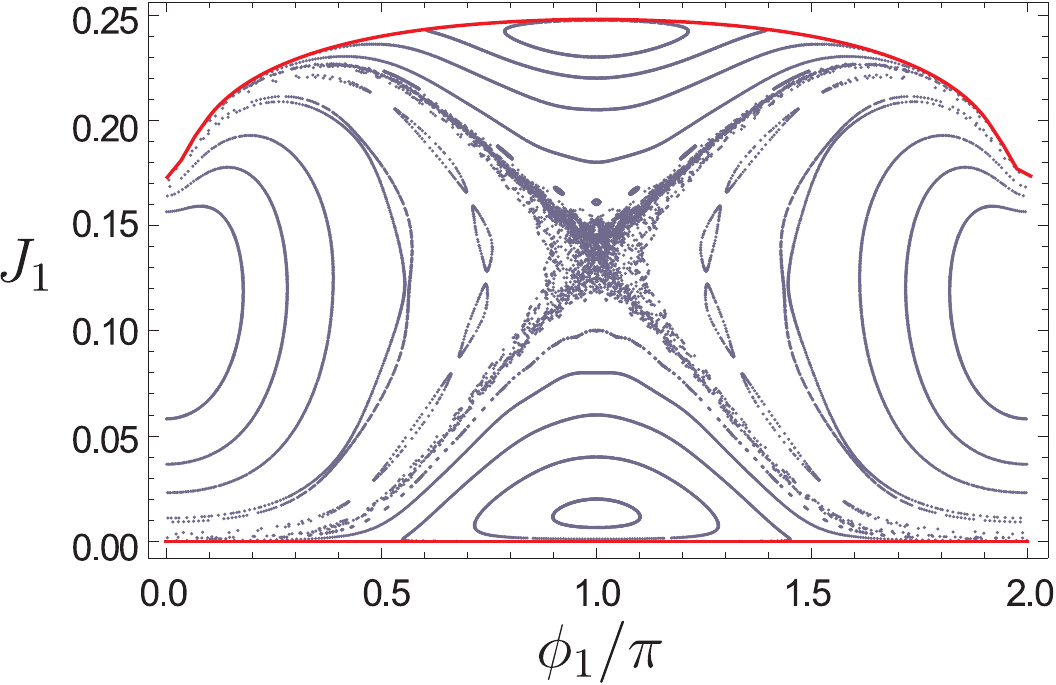}\\[10pt]
\hspace{-1cm}	$L=-0.02$&$L=0.0$&$L=0.05$\\
\hspace{-1cm}\includegraphics[width=6cm]{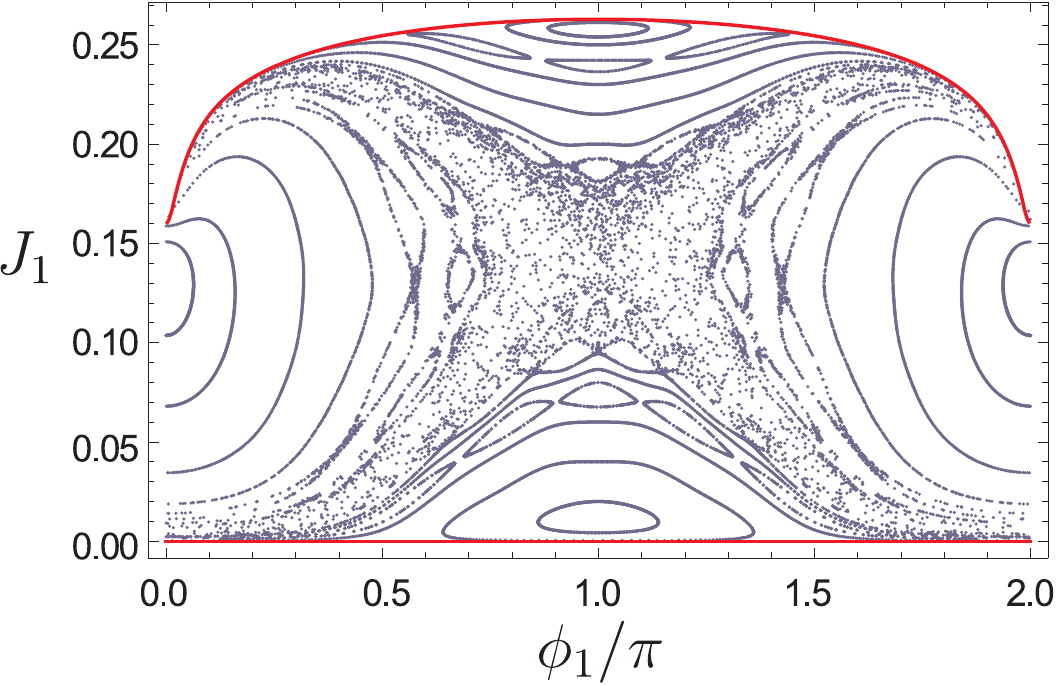}&\includegraphics[width=6cm]{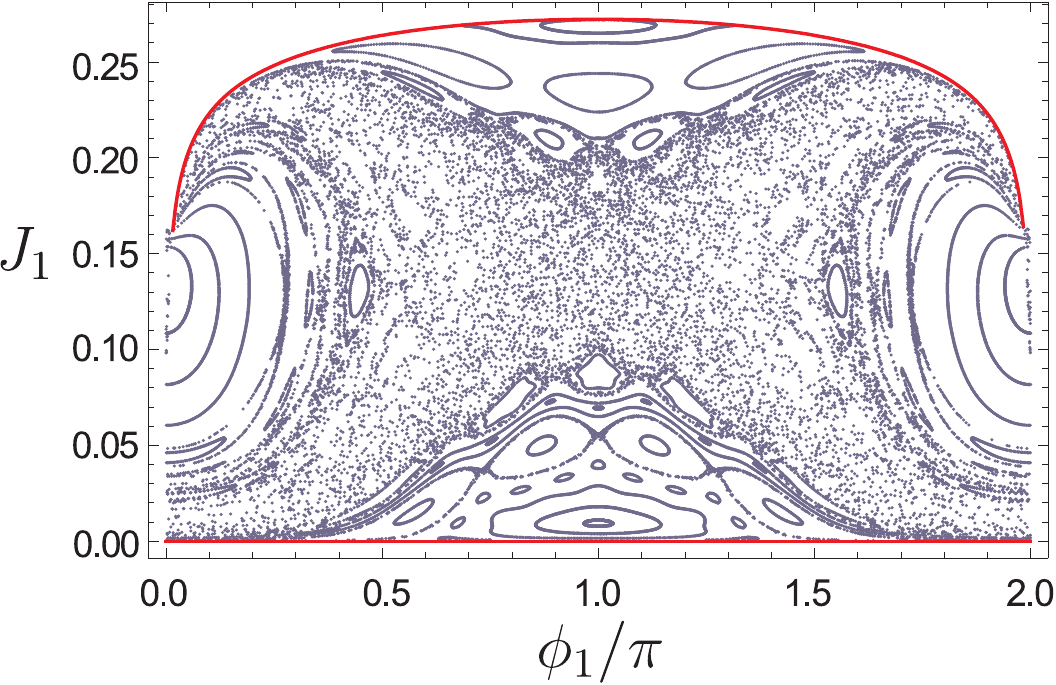}&\includegraphics[width=6cm]{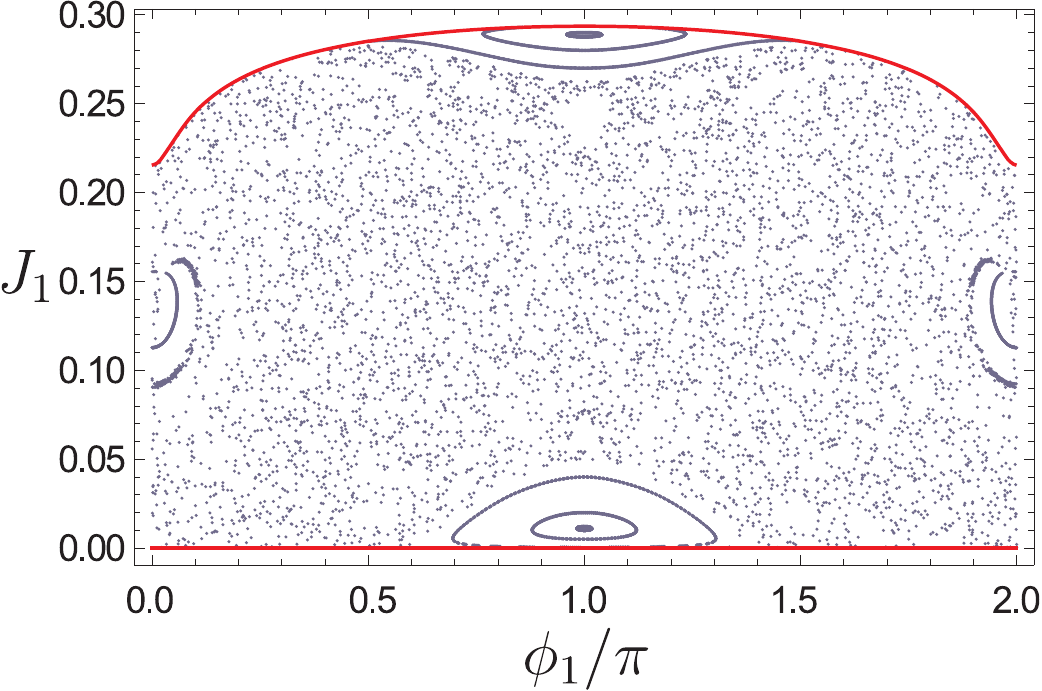}\\[10pt]
\hspace{-1cm}	$L=0.1$&$L=0.15$&$L=0.2$\\
\hspace{-1cm}	\includegraphics[width=6cm]{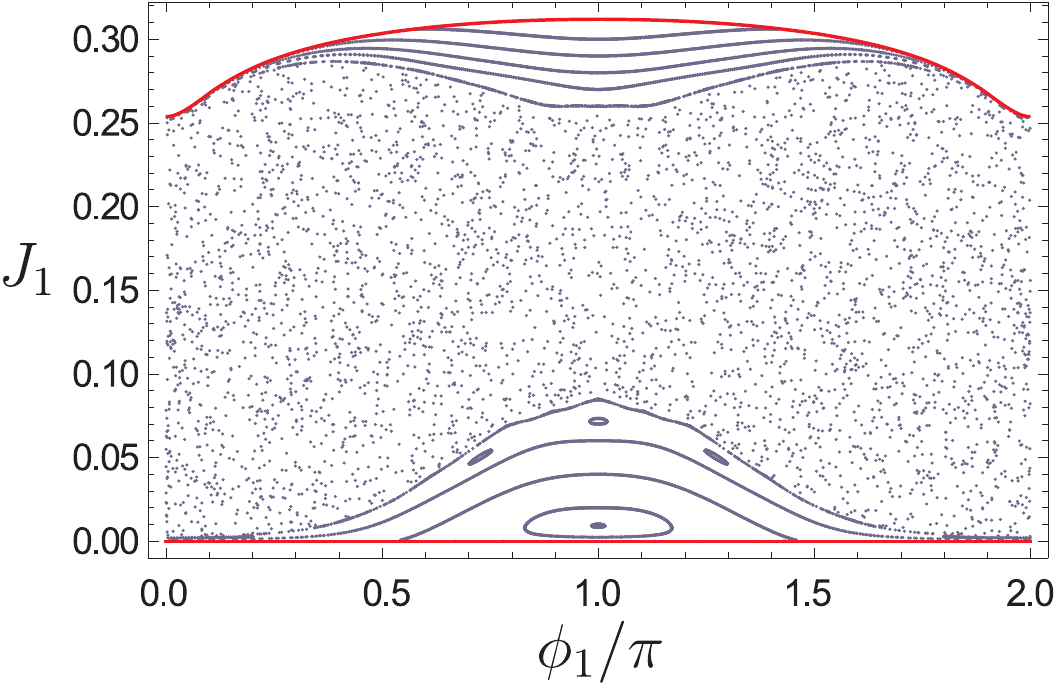}&\includegraphics[width=6cm]{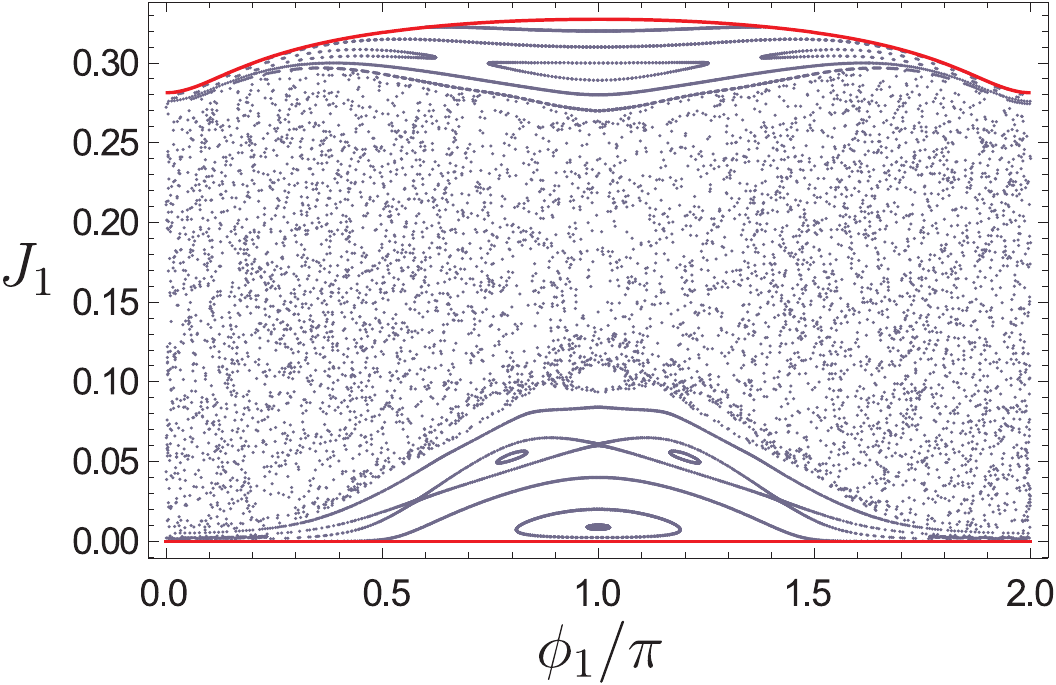}&\includegraphics[width=6cm]{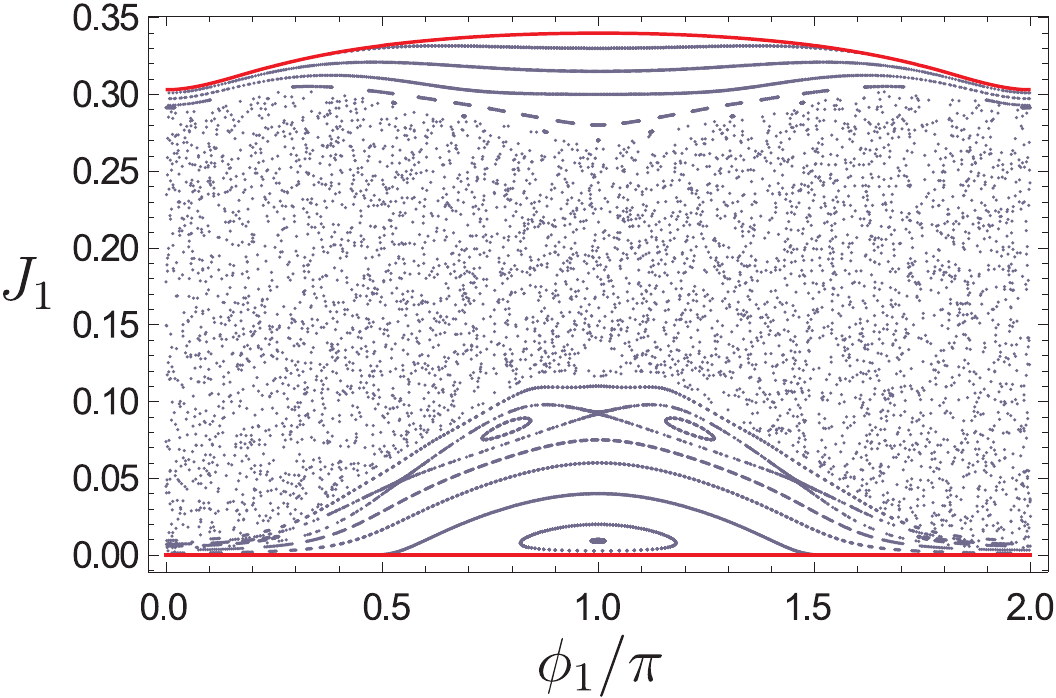}
	\end{tabular}
	\caption{(Color online) The graphs show the Poincar\'e section of the tripole
system in the same format as e.g. in Fig.~\ref{fig:sec_L_m_0_25} but 
for the case of progressively increasing $L$ in the range $-0.09\lesq L\lesq 0.2$. As can be observed,
the small chaotic region around the unstable central periodic orbit 
increases and finally spreads to almost all the phase space as the 
value of $L$ is increased, while ordered dynamics becomes confined into two regions in the top and bottom of the section.}
	\label{fig:chaos_rules}
\end{figure}

As the value of $L$ is increasing, we can see in Fig.~\ref{fig:chaos_rules} 
that the saddle point associated with the 
unstable periodic orbit is replaced by a small chaotic region 
around it, which becomes wider as $L$ becomes larger. I.e., the chaos,
as may be expected~\cite{lieber} ``emerges'' from this saddle point and 
gradually expands therearound. 
In addition we can observe the splitting of some of the invariant curves to 
form Poincar\'e-Birkhoff chains~\cite{lieber} of islands (e.g. for $L\geqslant-0.05$). In the center of these islands there are 
periodic orbits, which correspond to asymmetric configurations i.e.\, configurations with non-collinear initial conditions. Moreover, we observe the existence of islands of regularity inside the chaotic region, e.g. for $L=0$. At this point we would like to stress that in our study the classification of the trajectories as regular or chaotic seems more meaningful from a physical point, since periodic trajectories are isolated and their periodicity refers only to the reduced model. Actually, as we mentioned before, stable periodic trajectories correspond to regular quasi-periodic orbits in the ($x-y$) plane for the vortices. 

Nevertheless, the partition of ordered vs. chaotic regions on the sections can be clearly
discerned within the increasing $L$ diagrams of Fig.~\ref{fig:chaos_rules}. 
The chaotic region originating from the saddle point gradually expands along the
direction of the former stable and unstable manifold of the saddle,
overtaking the plane of the Poincar\'e section and gradually increasingly
restricting the ordered regions thereof. The latter become progressively
confined around the top and bottom periodic and the collision orbit of
the outer (left and right) regions of this cylindrical space.

\begin{figure}[ht]
	\centering	\includegraphics[width=7cm]{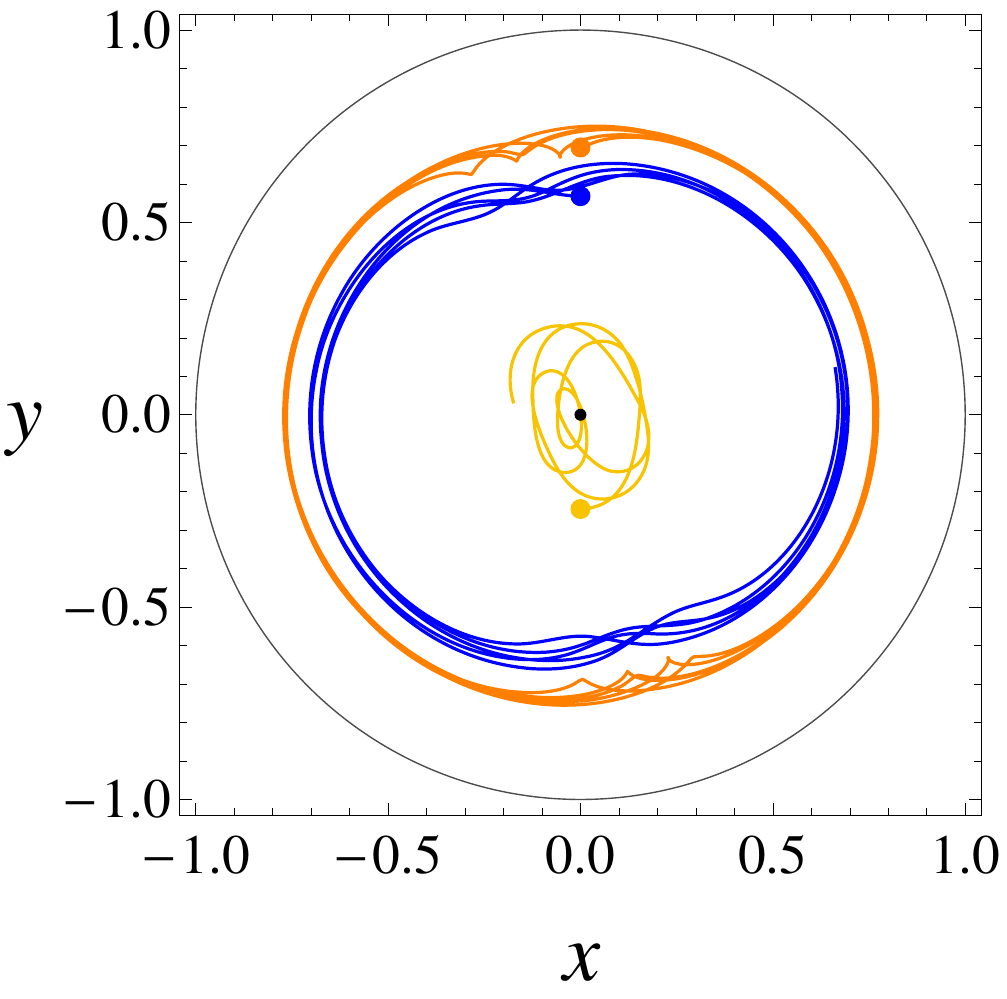}\hspace{0.5cm}\includegraphics[width=7cm]{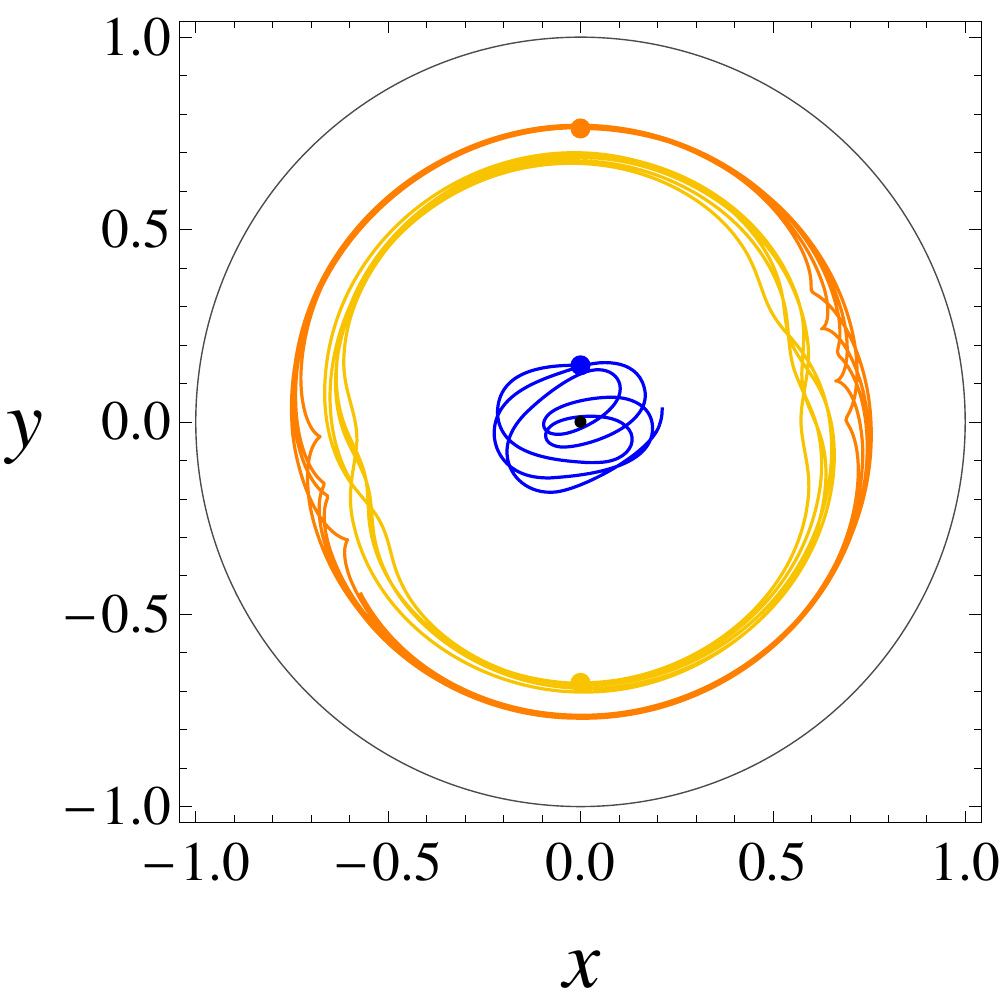}
	\caption{(Color online) The form of two regular orbits in the ($x-y$) plane for $L=-0.05$ and $t=15$. In the left panel, the orbit which corresponds to $(\phi_1, J_1)\simeq(\pi, 0.03)$ is depicted where the orbit of $S_3$ approaches the one of $S_2$. In the right 
panel, the $(\phi_1, J_1)\simeq(\pi, 0.23)$ is shown, where $S_1$ approaches $S_2$.}
	\label{fig:quasi_po_L_m_0_05}
\end{figure}

As the value of $L$ increases, the orbit of either $S_1$ or $S_3$ (depending on the initial conditions and consequently on the particular point on the section) approaches the one of $S_2$. For example, in Fig.~\ref{fig:quasi_po_L_m_0_05} we can see two regular orbits with symmetric configurations for $L=-0.05$ in the top
panels. The first one corresponds to $(\phi_1, J_1)\simeq(\pi, 0.03)$ and it is 
close to the lower symmetric stable periodic orbit while the other corresponds to $(\phi_1, J_1)\simeq(\pi, 0.23)$ and it is close to the upper symmetric stable periodic orbit as can be seen in Fig.~\ref{fig:chaos_rules}. In the first figure we can see how the interaction of $S_2$ and  $S_1$ affects their orbits, while in the second we can observe the same feature for the $S_2$ and $S_3$ vortices.

\subsection{Dynamics for $L>0$}

\begin{figure}[h]
	\centering
		\includegraphics[width=7cm]{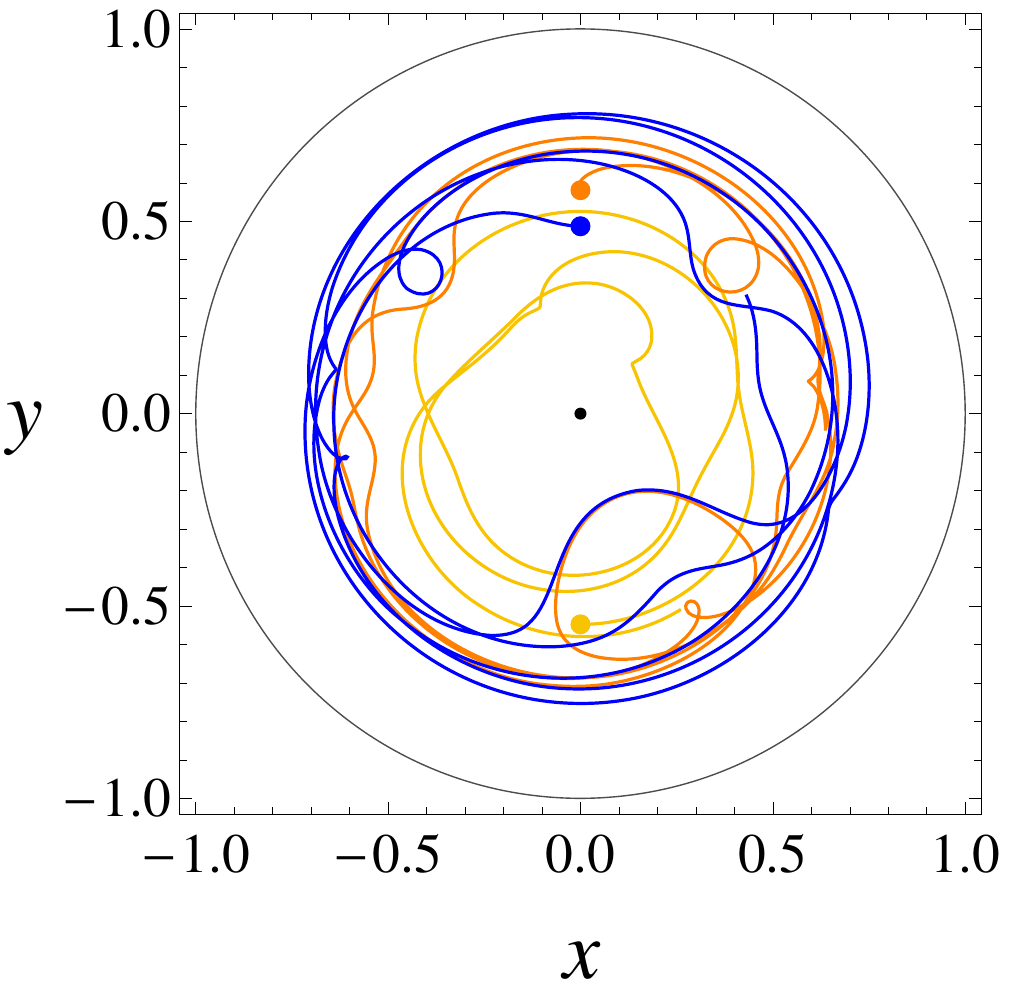}\hspace{0.5cm}\includegraphics[width=7cm]{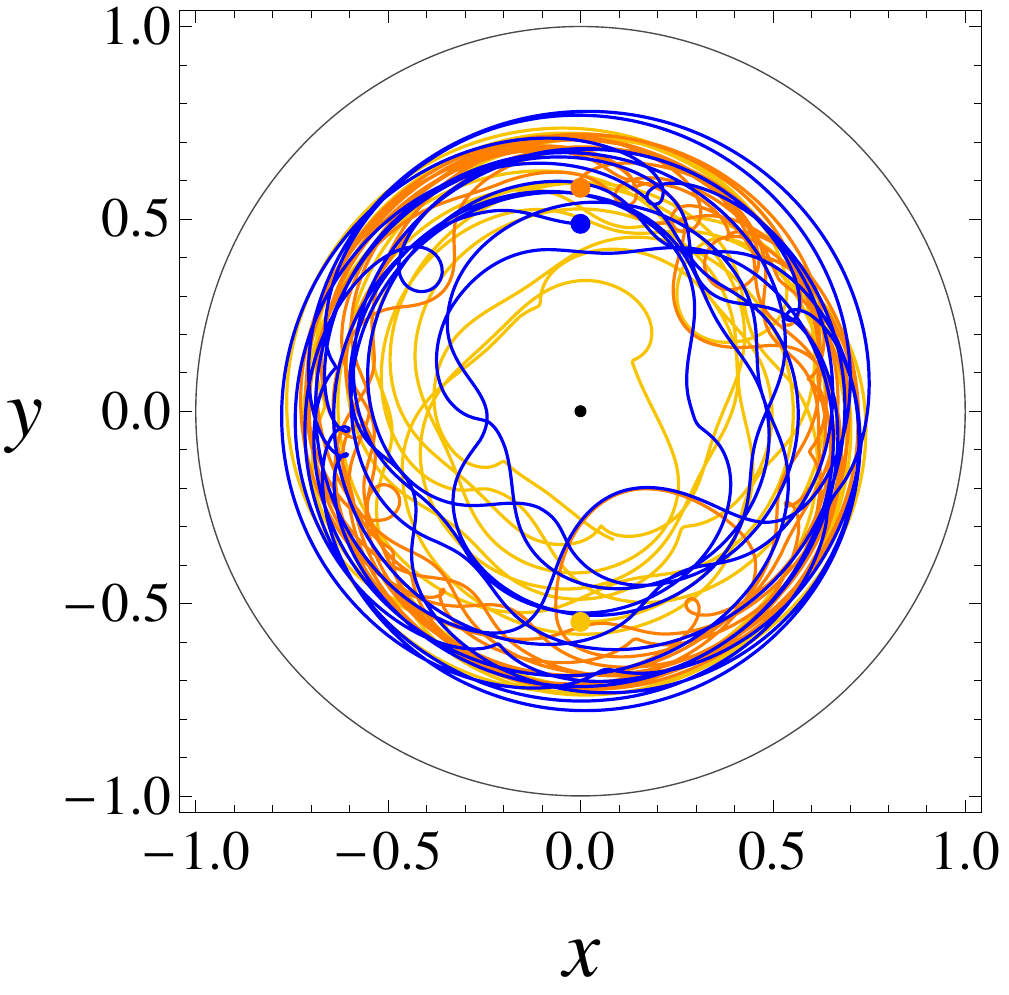}
	\caption{(Color online) A chaotic mixing orbit for $L=0.1$ for $(\phi_1, J_1)=(\pi, 0.15)$. In the left panel, the orbit has evolved for $t=15$, while in the right one for $t=50$.} 
	\label{fig:chaotic_o_L_0_1}
\end{figure}

As we progress into the predominantly chaotic phase space associated
with positive values of $L$, 
the strong interaction between the vortices is responsible for the expansion 
of the chaotic region. The distinction of the motion in the ($x-y$) plane of 
an orbit which corresponds to a chaotic evolution in the Poincar\'e section is very clear from the one of the ordered orbits shown previously in the
($x-y$) plane. For example, in Fig.~\ref{fig:chaotic_o_L_0_1} the motion which 
corresponds to a chaotic orbit for $L=0.1$ is shown. In this case we can see 
how the orbits of the three vortices are mixing and filling the 
plane as time evolves, in contrast with the regular 
ones where the vortices occupy distinct regions in the ($x-y$) plane. 

\begin{figure}[h]
	\centering
		\includegraphics[width=10cm]{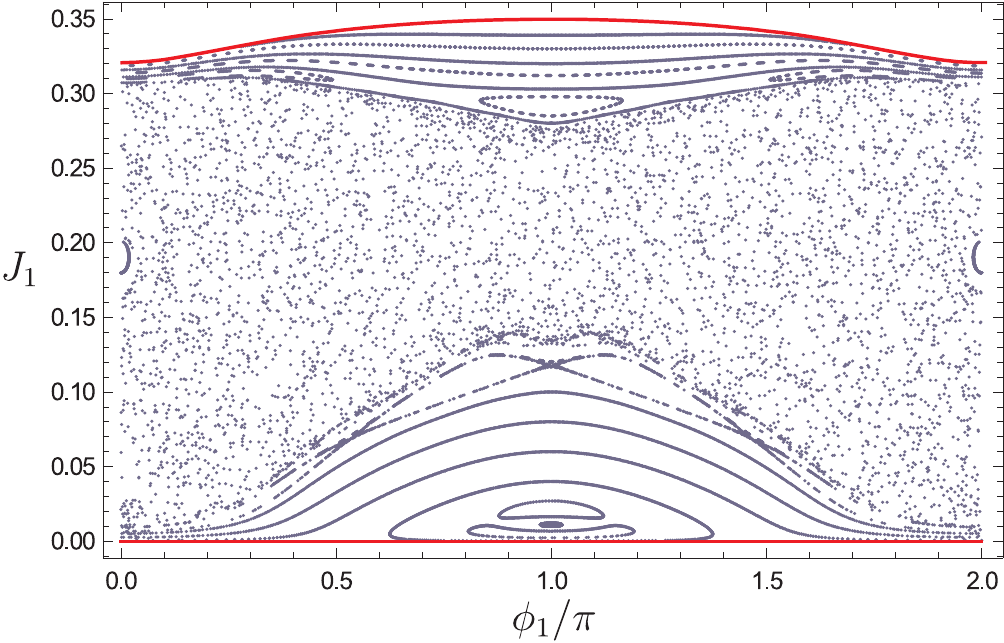}
	\caption{(Color online) The Poincar\'e section for $L=0.25$. In this area of values of $L$, the two persisting regular regions of the section have occupied their
maximal area before shrinking again in size for larger $L$.} 
	\label{fig:pnc_L_0_25}
\end{figure}

Although, the chaotic region grows larger and tends to occupy the whole phase-space for increasing values of $L$, there are two regular regions around $\phi_1=\pi$ which persist and apparently expand for a range of positive $L$'s beyond $L>0.05$. These 
regions acquire their maximal size (i.e., fraction of the plane's area) close to the value of $L=0.25$ as can be seen in Fig.~\ref{fig:pnc_L_0_25},
before shrinking again for larger values of $L$. Such features will be
discussed in a more quantitative fashion in a forthcoming work~\cite{kyr}. 

\begin{figure}[htbp]
	\begin{tabular}{cc}
$L=0.244$&$L=0.245$\\	
\includegraphics[width=6cm]{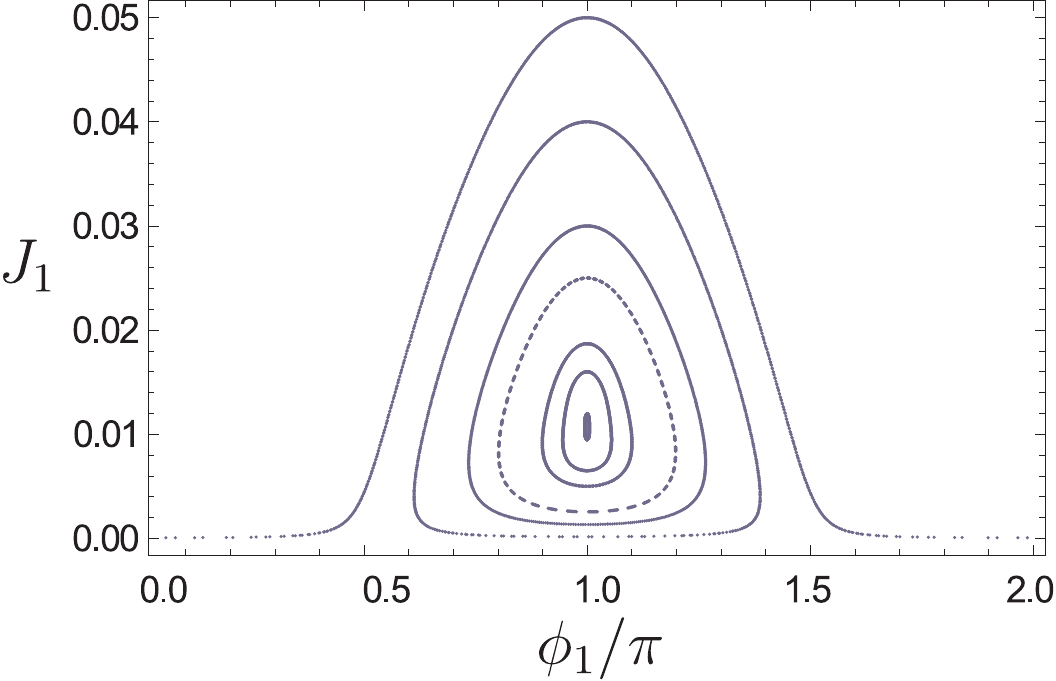}&\includegraphics[width=6cm]{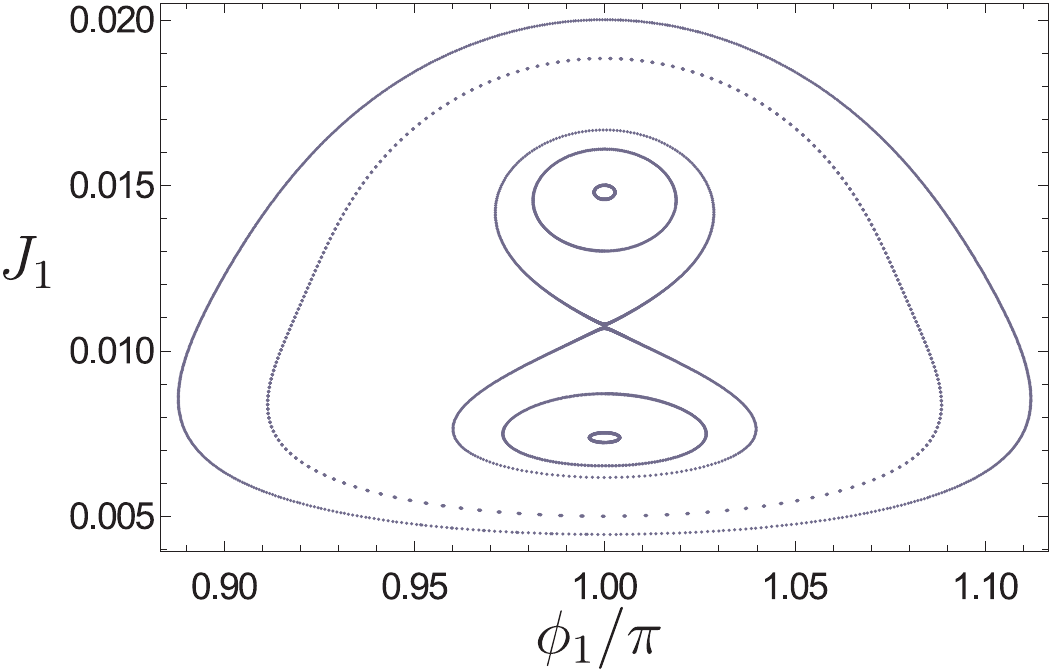}\\[10pt]
$L=0.246$&$L=0.247$\\[2pt]	\includegraphics[width=6cm]{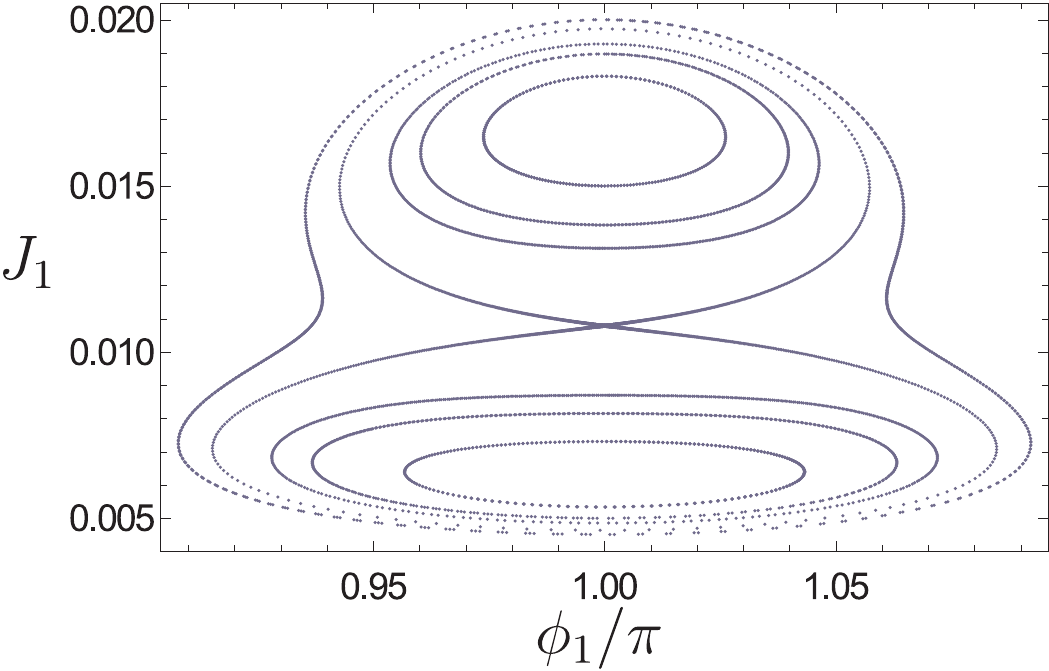}&\includegraphics[width=6cm]{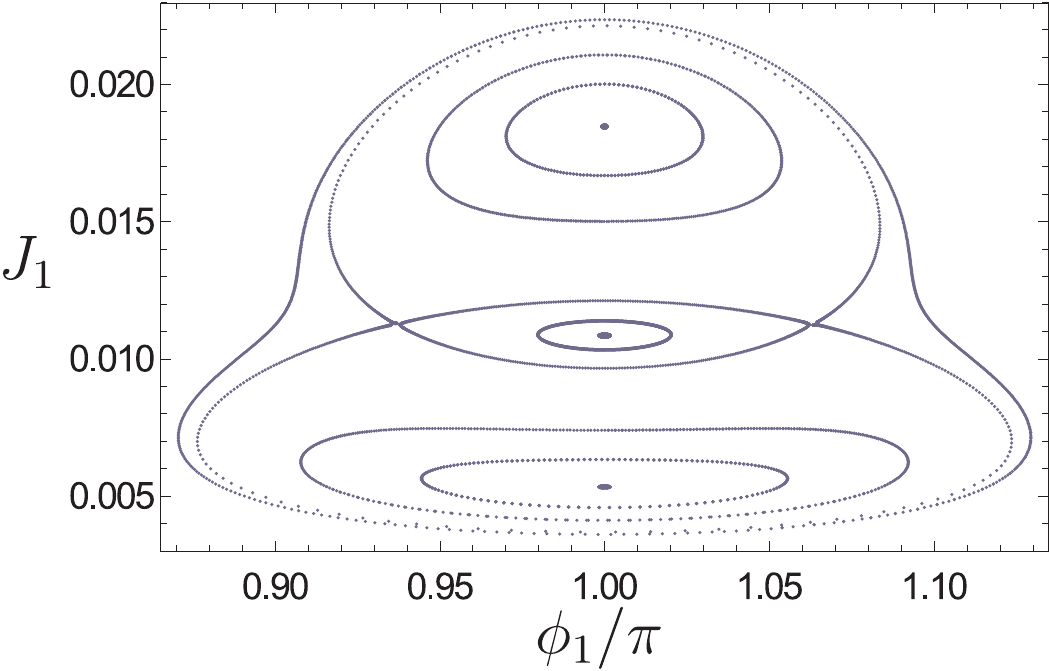}
	\end{tabular}
	\caption{(Color online) A magnification of the Poincar\'e sections
is shown for the period doubling scenario described in the text
and occurring for $0.2<L<0.247$.}
	\label{fig:period_doubling}
\end{figure}

\begin{figure}[htbp]
	\begin{tabular}{ccc}
$L=0.244$&$L=0.245$\\
\hspace{0.5cm}\includegraphics[width=5cm]{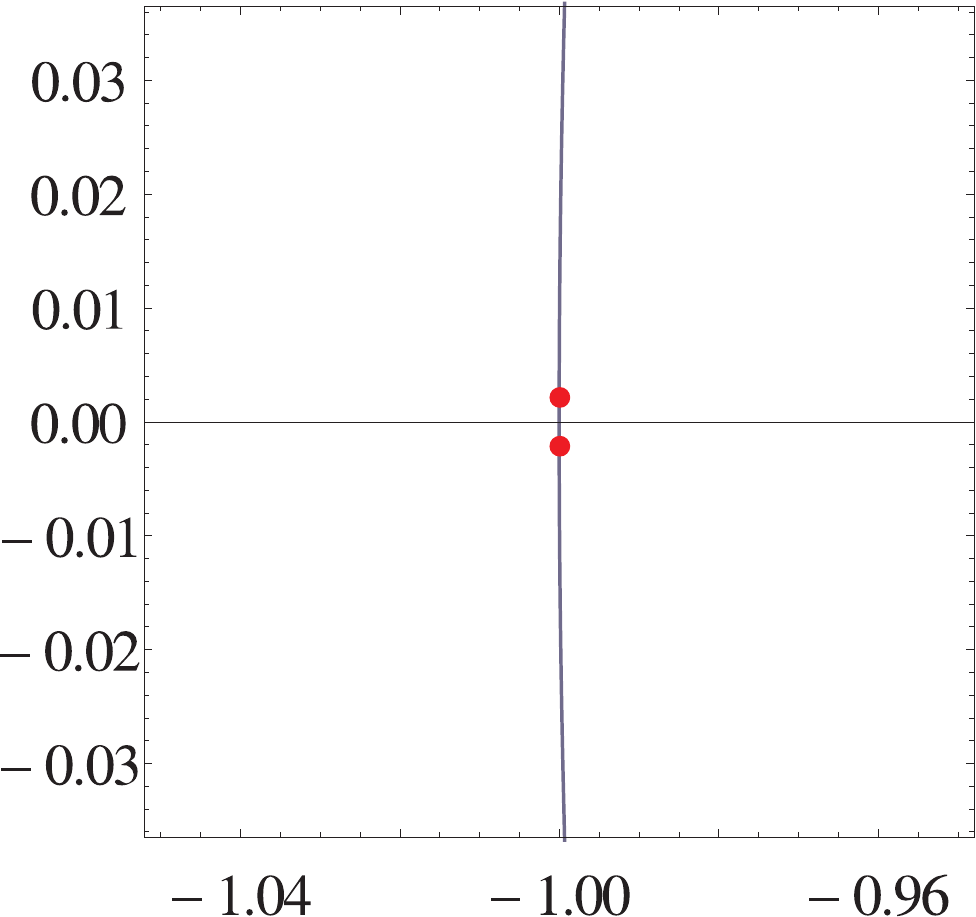}&\hspace{0.5cm}\includegraphics[width=5cm]{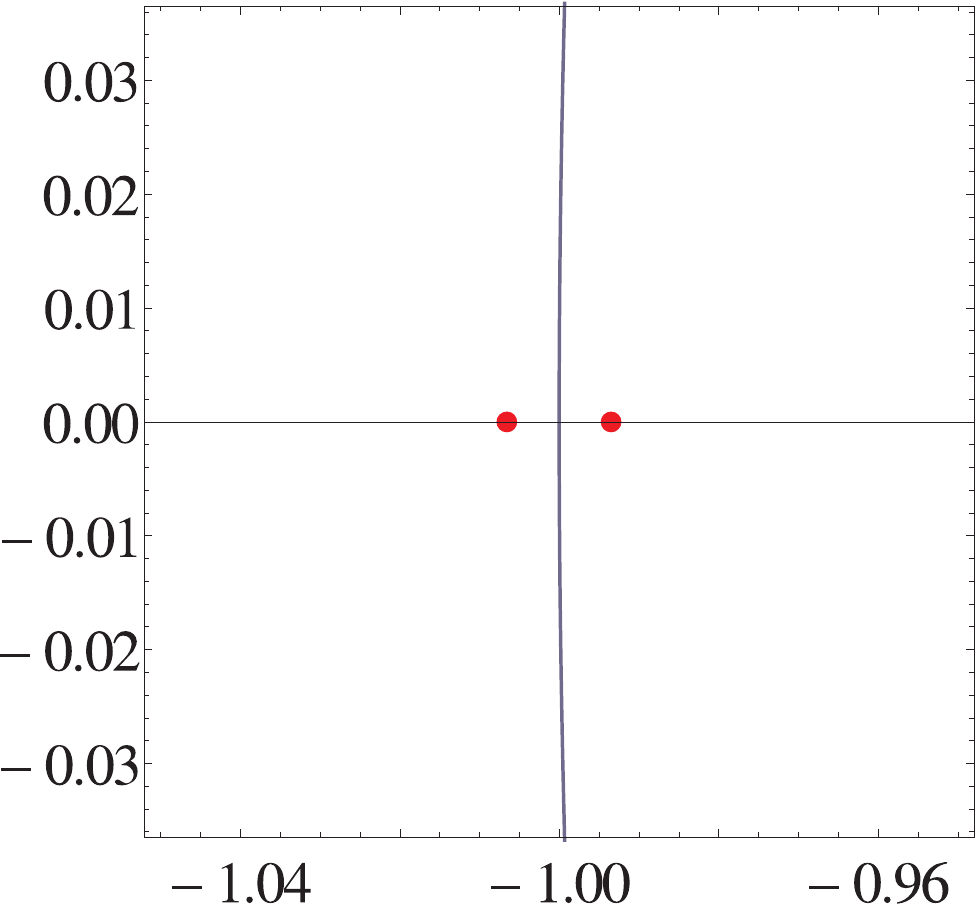}\\[15pt]
$L=0.246$&$L=0.247$\\	\includegraphics[width=5cm]{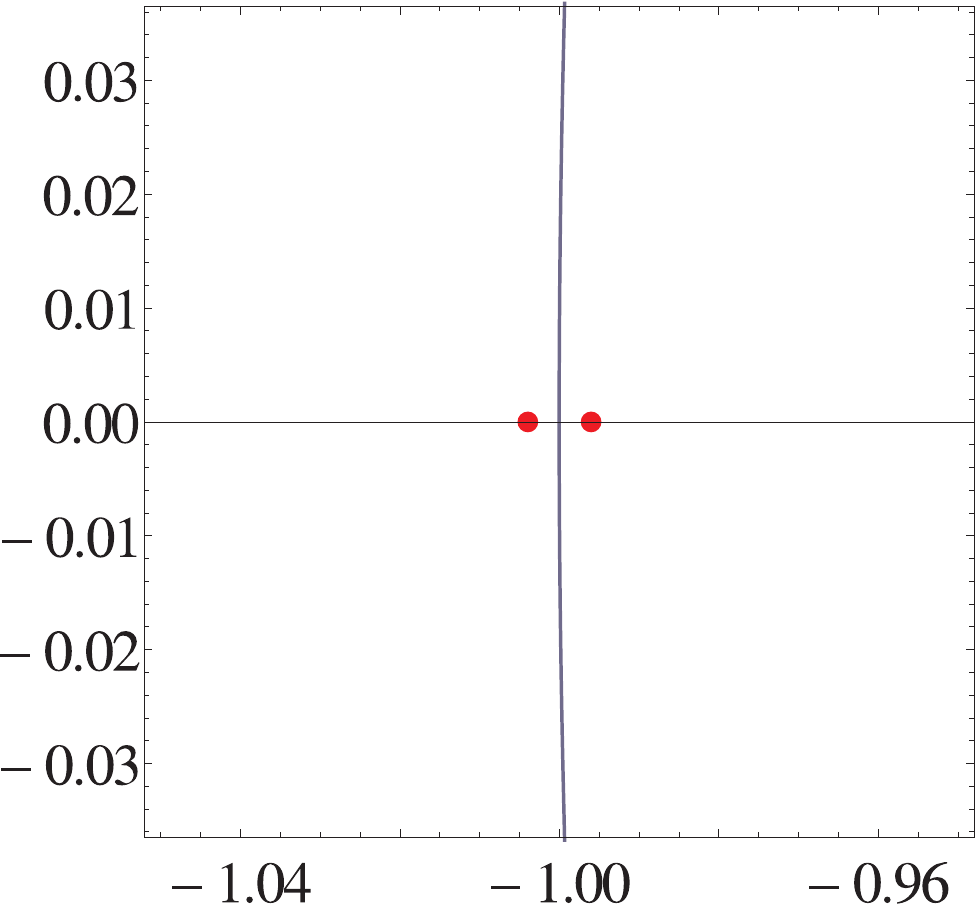}&\hspace{0.5cm}\includegraphics[width=5cm]{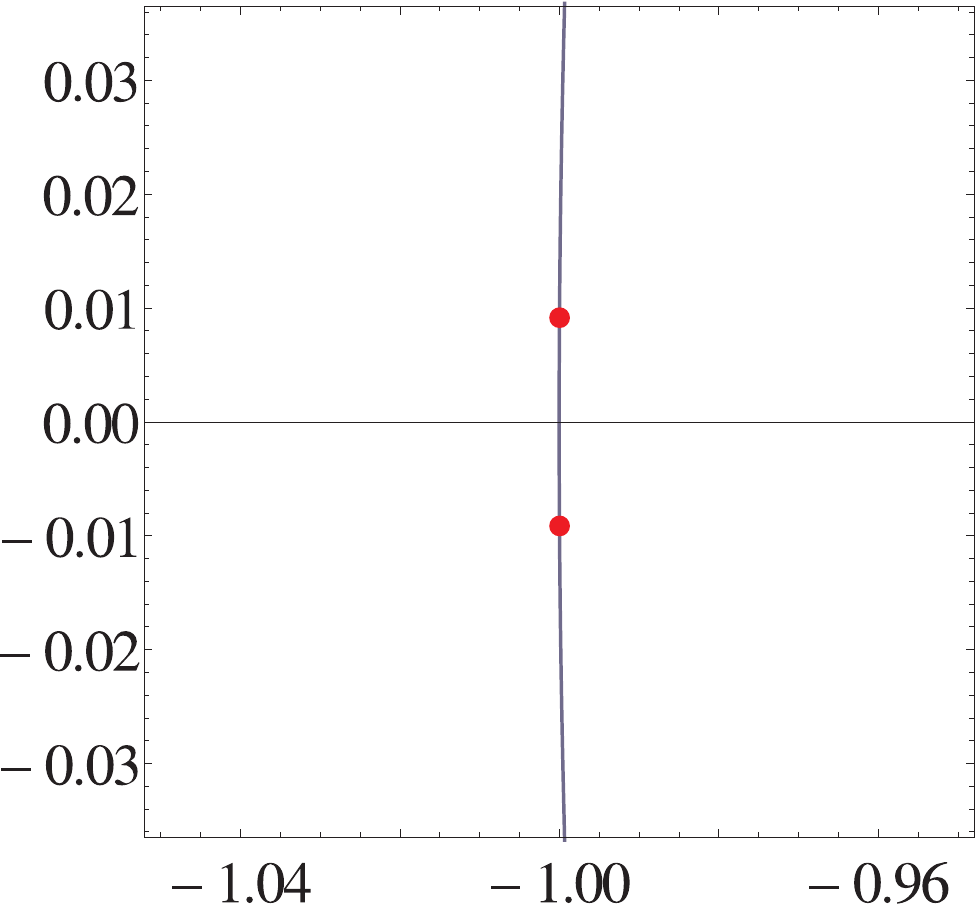}\\
	\end{tabular}
	\centering
	\hspace{-0.5cm}\includegraphics[width=7cm]{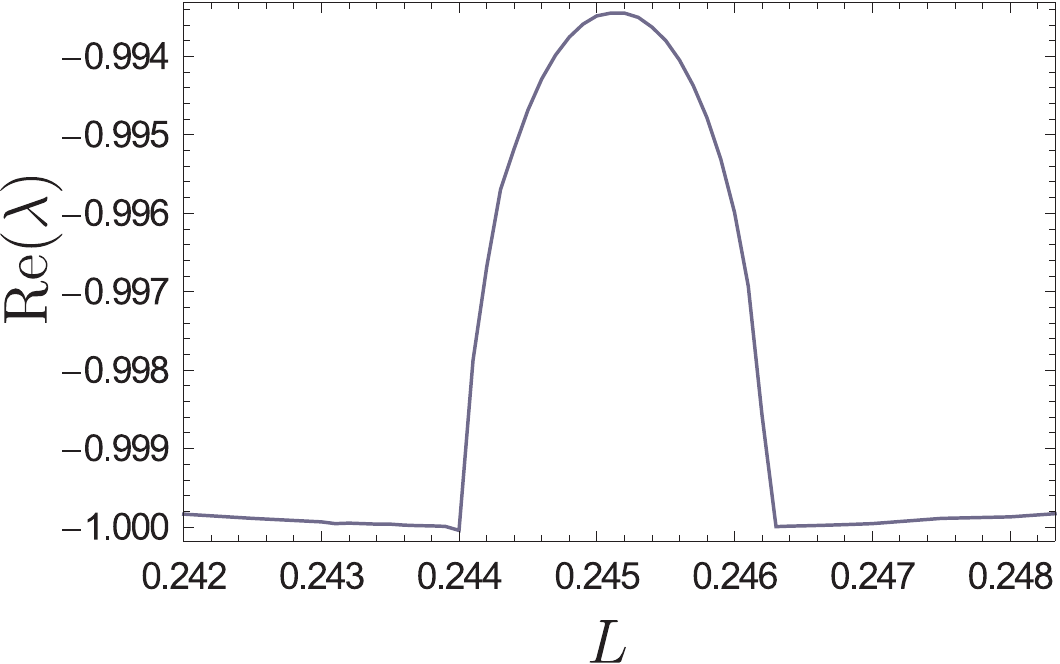}\hspace{0.5cm}\includegraphics[width=7cm]{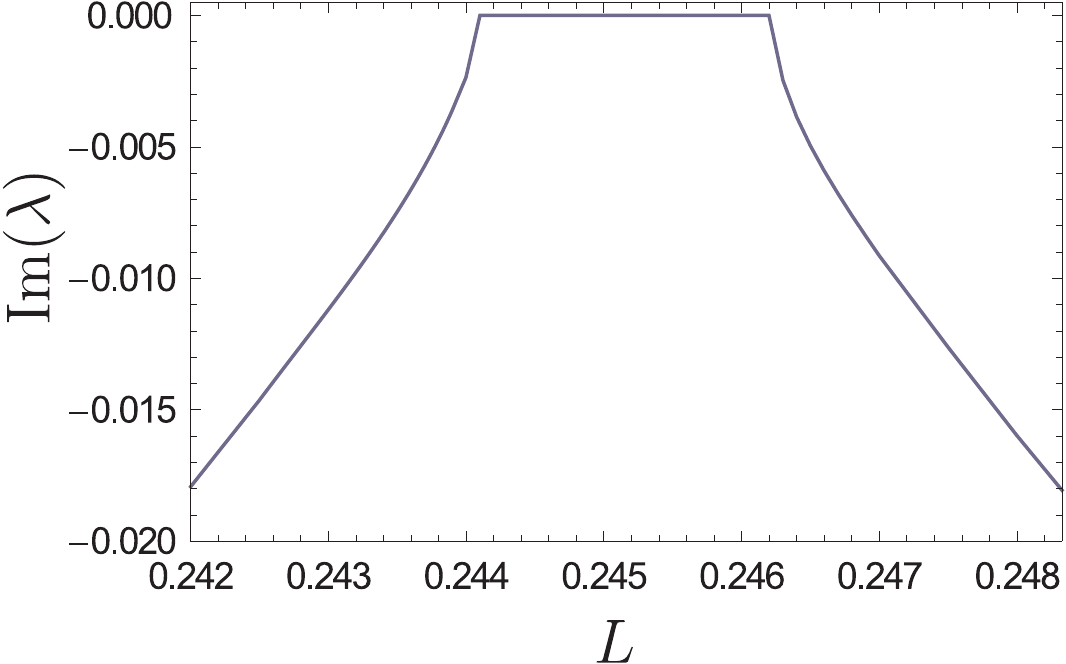}
	\caption{(Color online) Top two rows: the dependence of the relevant
Floquet multipliers of the destabilizing and restabilizing 
periodic orbit associated with the period doubling scenario described
in the text.
Bottom row: the real and imaginary parts of the associated multipliers.}
	\label{fig:eig_period_doubling}
\end{figure}

In addition to the expansion of the chaotic region, within
the region $0.2<L<0.247$, an interesting bifurcation scenario unfolds. 
At $L\simeq0.2441$, a period-doubling bifurcation of the stable symmetric periodic orbit which corresponds to $(\phi_1, J_1)\simeq(\pi, 0.011)$ occurs and the stable orbit is replaced by an unstable (saddle point) one and an emerging
 stable one of double the original period. However, at $L\simeq0.2463$ an inverse period doubling bifurcation occurs, whereby the unstable periodic orbit becomes stable again giving rise to an extra asymmetric periodic orbit, which has double the period with respect to the original one.  
This way we end up with two stable symmetric periodic orbits one with double the period of the other and a doubled-period asymmetric unstable periodic orbit. The whole scenario can be seen in the panels of Fig.~\ref{fig:period_doubling}, while in those of Fig.~\ref{fig:eig_period_doubling} the corresponding eigenvalues of the Floquet matrix of the central periodic orbit are shown in the vincinity of $-1$, clearly illustrating the forward and reverse period-doubling
bifurcation. 

\begin{figure}[htbp]
	\begin{tabular}{cc}
	$L=0.3025$&$L=0.303$\\[10pt]
	\includegraphics[height=5cm]{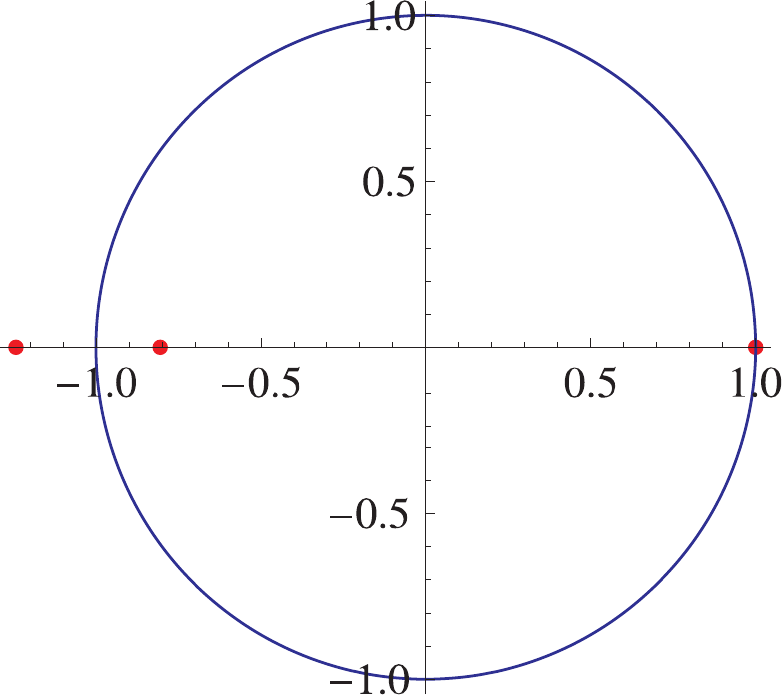}&\hspace{0.5cm}\includegraphics[width=5cm]{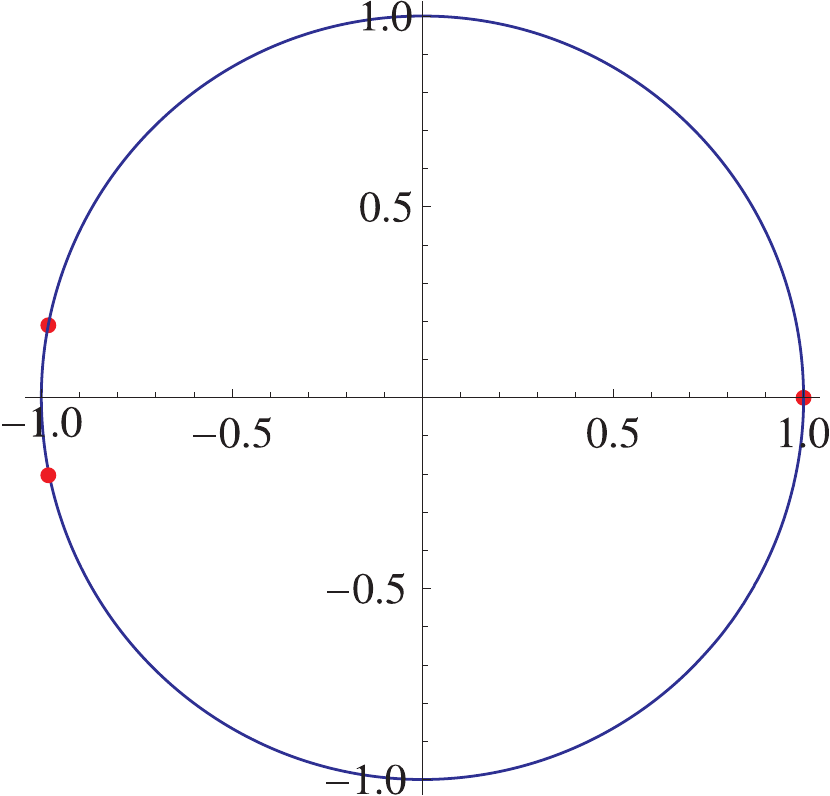}
		\end{tabular}
	\caption{(Color online) The Floquet multipliers 
for the inverse period doubling scenario of the asymmetric periodic orbit
which leads to its stabilization.}
	\label{fig:eig_inverse_period_doubling}
\end{figure}

\begin{figure}[htbp]
	\centering
		\includegraphics[height=5cm]{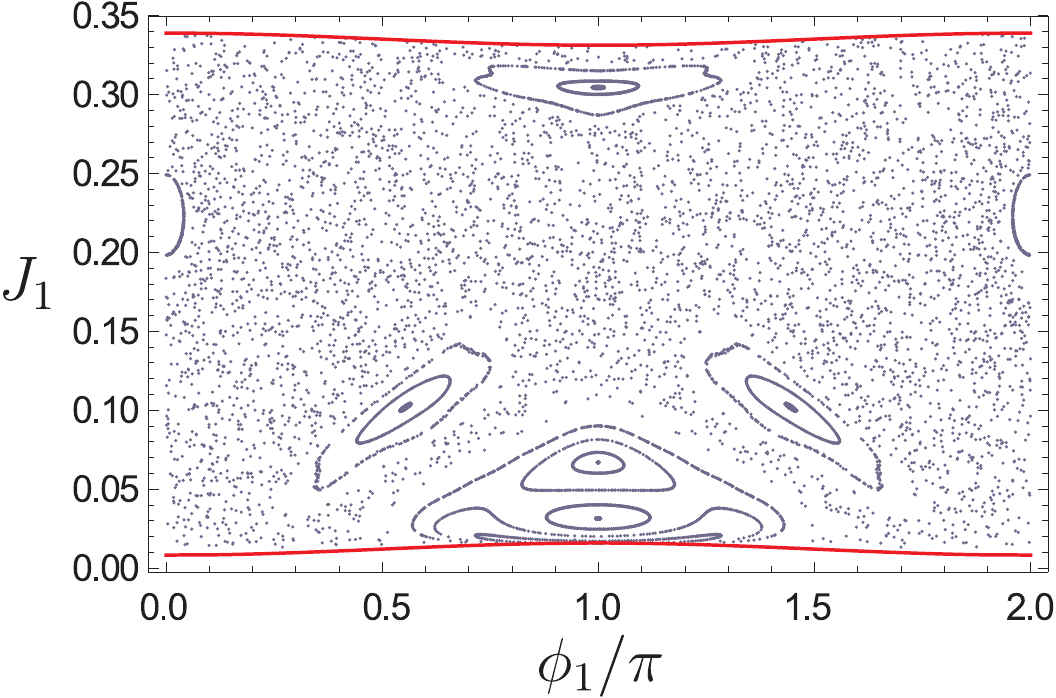}\hspace{1cm}\includegraphics[height=5cm]{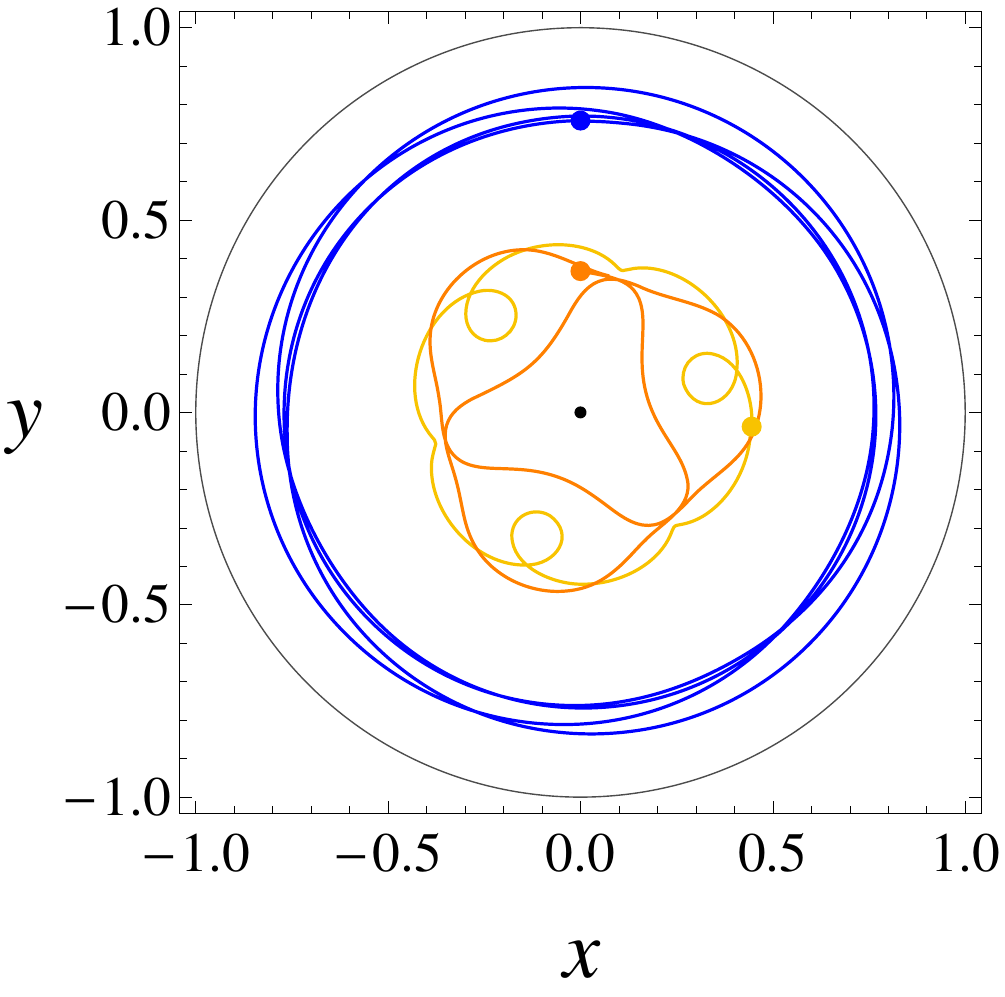}
	\caption{(Color online) Left panel: the Poincar\'e section for $L=0.35$. We can observe the two 
additional regions of regular orbits around the asymmetric double period periodic orbit. Right panel: the form of the asymmetric periodic orbits in the $(x-y)$ plane is shown.}
	\label{fig:L_0_35}
\end{figure}

For $L\simeq0.3028$ a further significant structural change occurs. An 
asymmetric stable periodic orbit at $(\phi_1, J_1)\simeq(0.505\pi, 0.101)$ 
appears through an inverse period doubling bifurcation.
This is shown in detail in Fig. \ref{fig:eig_inverse_period_doubling}
through the Floquet multipliers of the relevant periodic orbit. 
In the left panel of Fig.~\ref{fig:L_0_35} we can see  the Poincar\'e section of the system 
for $L=0.35$. We can distinguish the two additional regular regions around the asymmetric periodic orbit. Note that, from this value of $L$ onward, 
the upper boundary of the section is calculated by the $R_2=0$ requirement 
rather that the $R_3=0$ that was used up to this point. This happens because 
as the value of $L$ is increasing, $R_2$ becomes smaller. Consequently, the $S_2$ vortex becomes a candidate for crossing the center.
The right panel of Fig.~\ref{fig:L_0_35} shows 
the time evolution in the $(x-y)$ plane of the asymmetric periodic orbits for this value of $L$.

\begin{figure}[htbp]
	\centering
	\begin{tabular}{cc}
		$L=0.365$ & $L=0.37$\\
		\includegraphics[width=8cm]{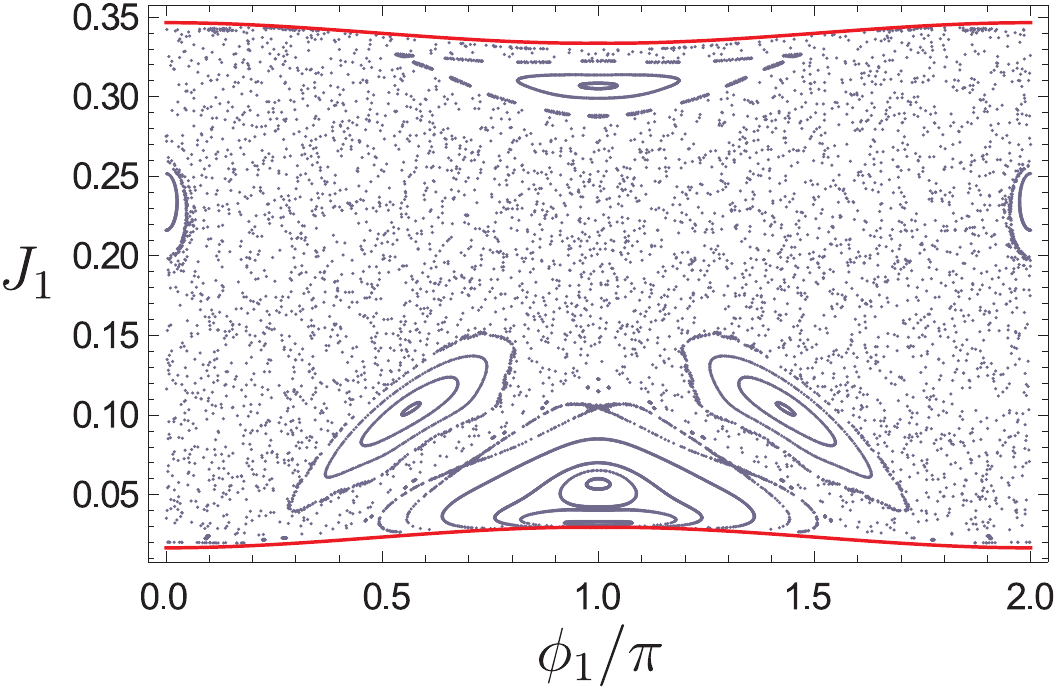}&\includegraphics[width=8cm]{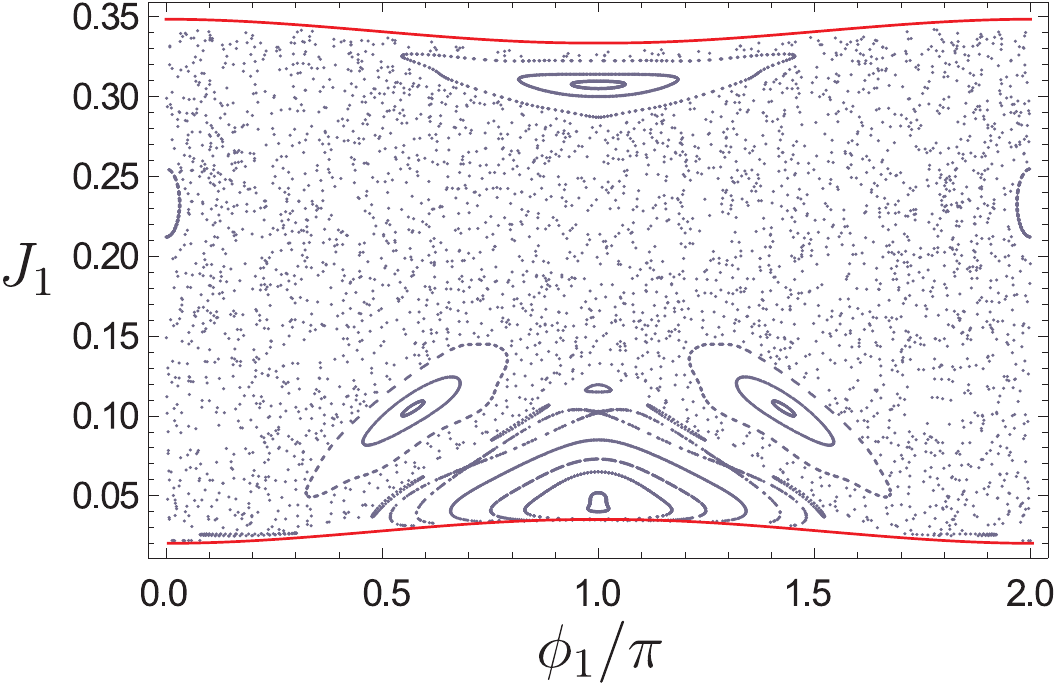}\\[10pt]
		$L=0.42$ & $L=0.43$\\
		\includegraphics[width=8cm]{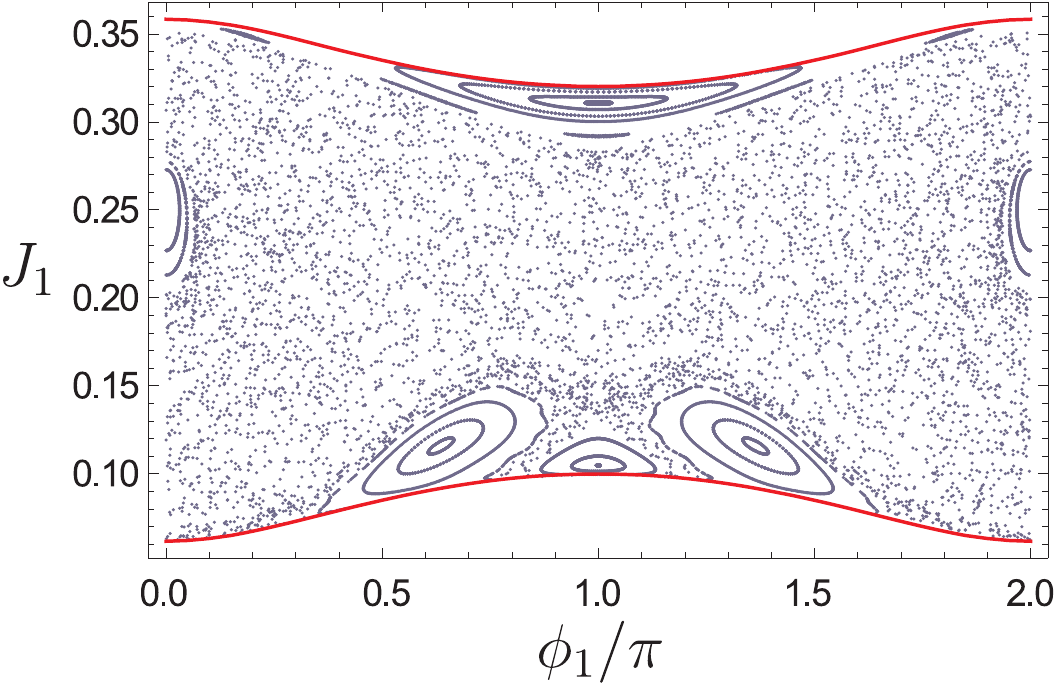}&\includegraphics[width=8cm]{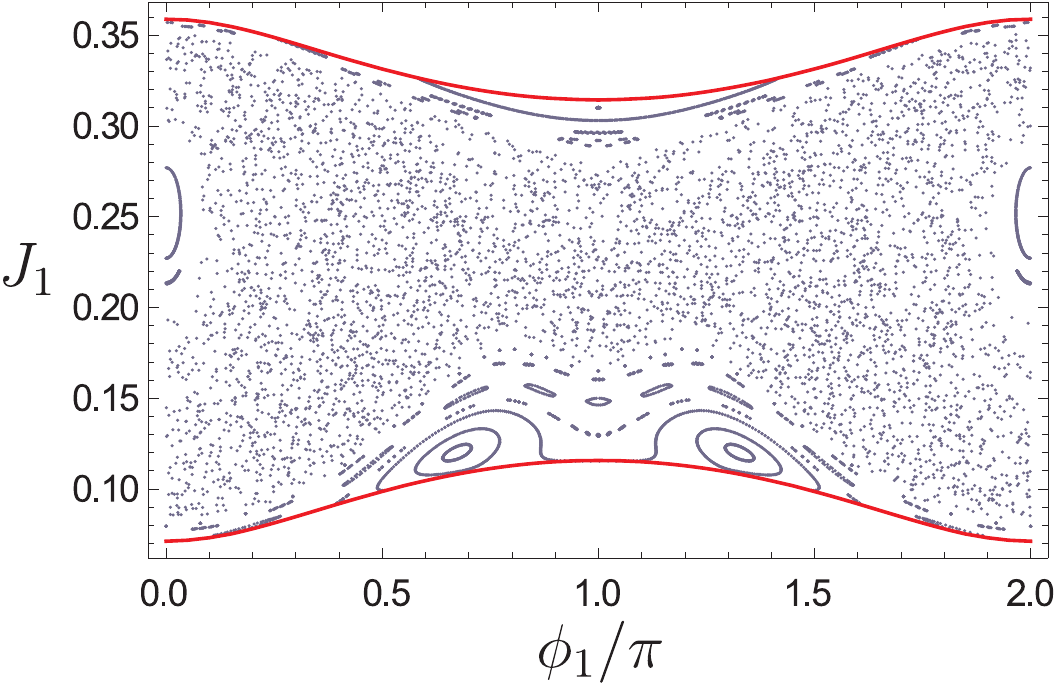}\\[10pt]	
		\end{tabular}
	\caption{(Color online) The Poincar\'e sections are shown in a similar 
way as in the earlier figures but now for $0.365\lesq L\lesq0.43$.}
	\label{fig:pnc_mL1}
\end{figure}

In Fig.~\ref{fig:pnc_mL1} the behavior of the system for $0.365 \lesq L\lesq 0.43$ is shown. 
In the section which corresponds to $L=0.365$ we can see that the upper and the asymmetric regions of stability become wider, while in the lower region of stability the inverse bifurcation scenario of this described in Fig.~\ref{fig:period_doubling} occurs, which leads the stable and unstable double-period and the stable single-period orbits to be replaced by a single-period orbit as it is shown in the surface of section for $L=0.37$. For higher values of $L$ we see that the area of the permitted orbits in the Poincar\'e section ``shrinks'', i.e.\ the two boundary-curves approach around $\phi_1=\pi$ due to the energy and angular momentum conservation constraints. At the same time, we see how the region around the lower symmetric periodic orbit is ``squeezed'' by the boundary, and finally for $L=0.43$ disappears.

\begin{figure}[htbp]
	\centering
	\begin{tabular}{cc}
		$L=0.44$ & $L=0.46$\\	
		\includegraphics[width=8cm]{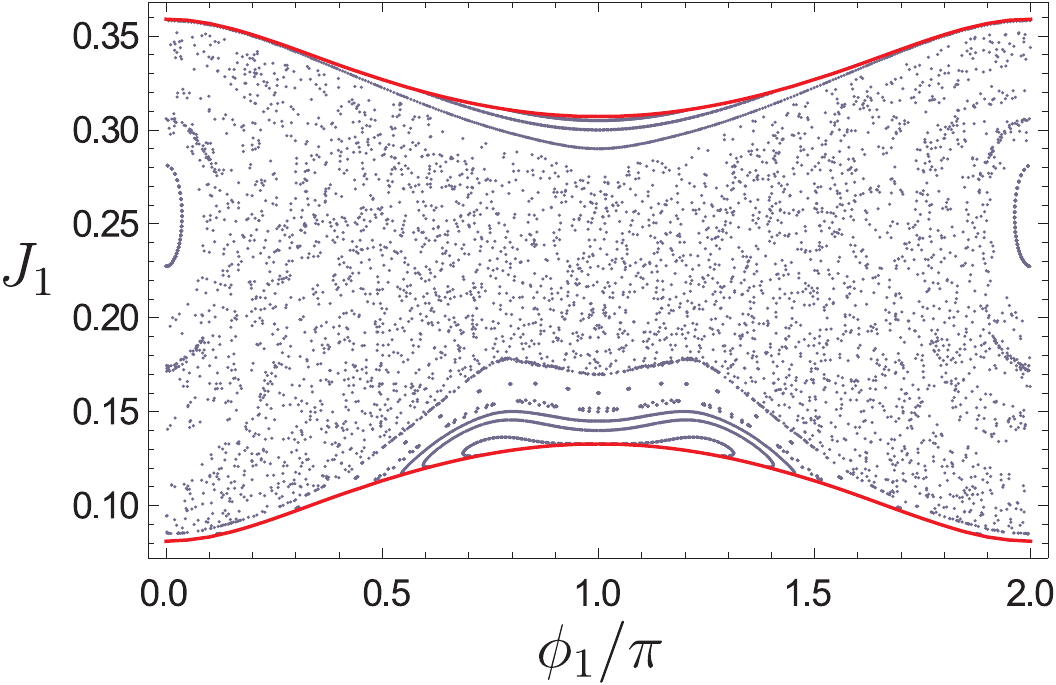}&\includegraphics[width=8cm]{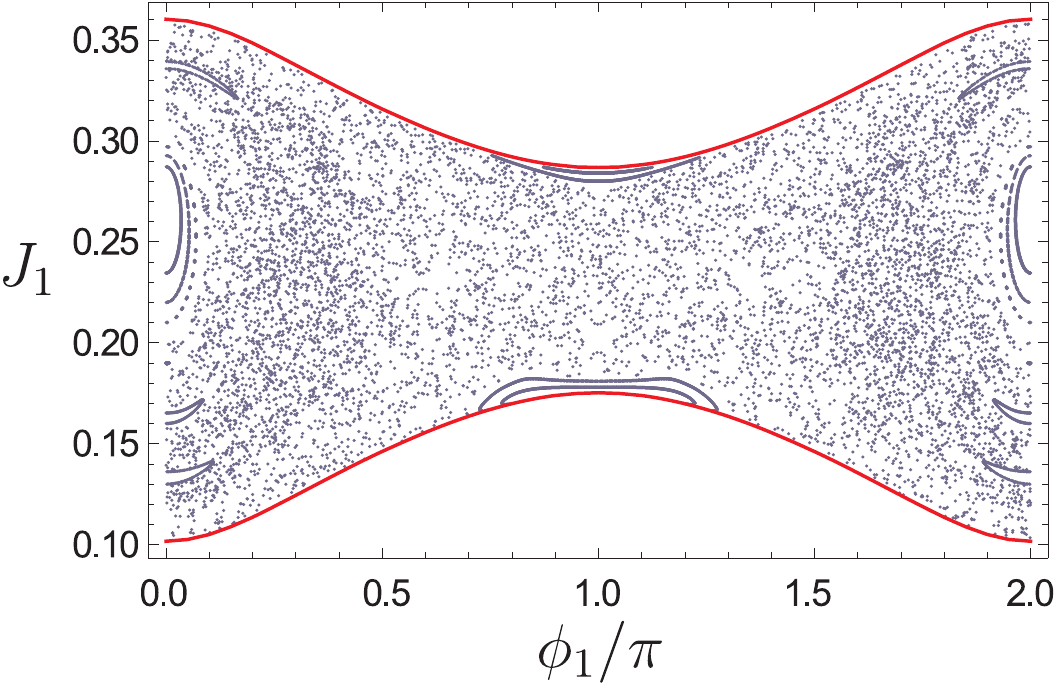}\\[10pt]
		$L=0.47$ & $L=0.473$\\	
		\includegraphics[width=8cm]{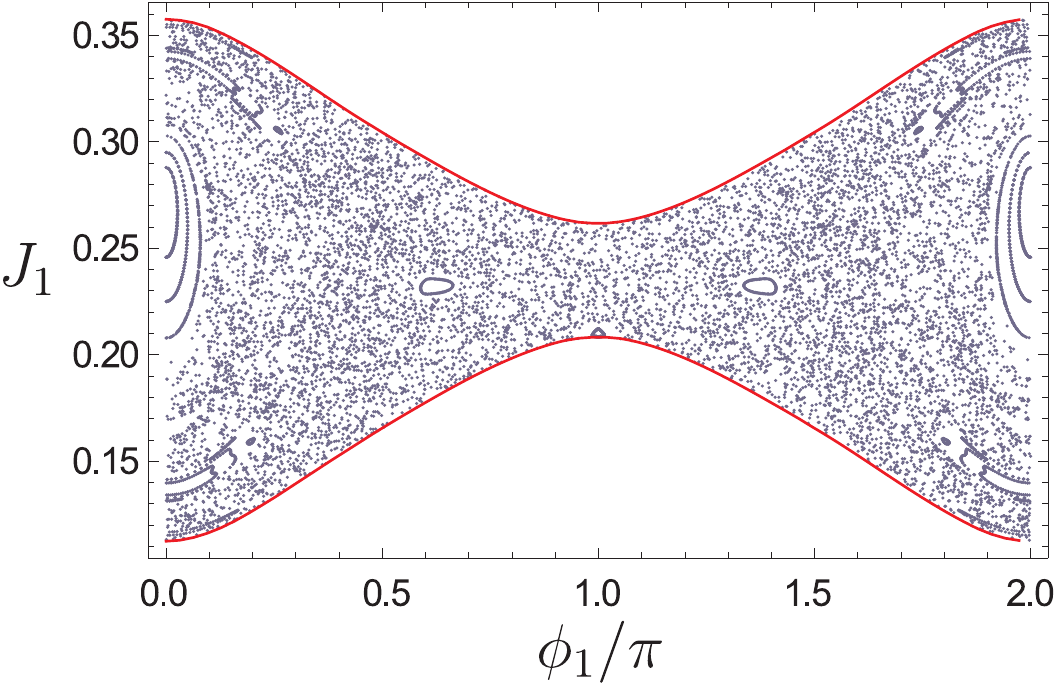}&\includegraphics[width=8cm]{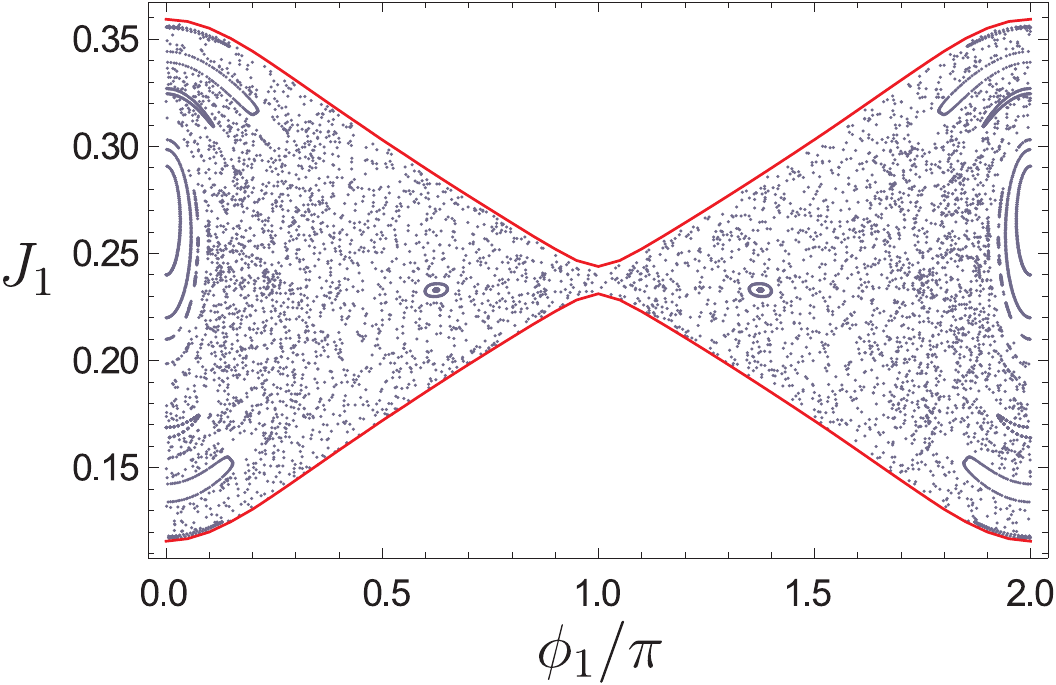}\\[10pt]
		$L=0.50$ & $L=0.55$\\
		\includegraphics[width=8cm]{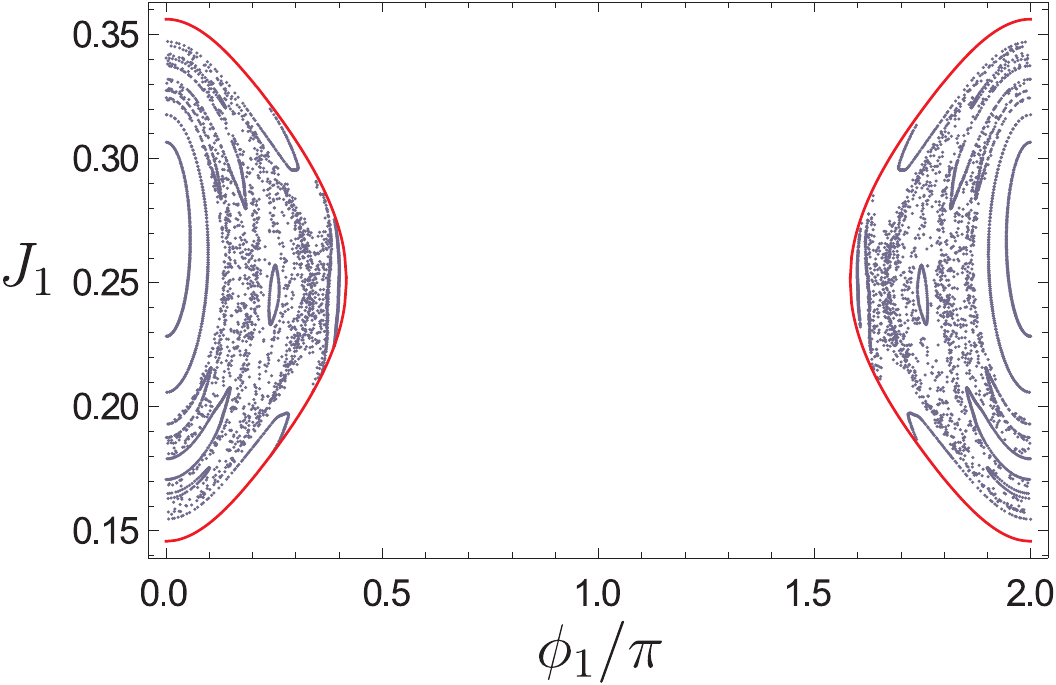}&\includegraphics[width=8cm]{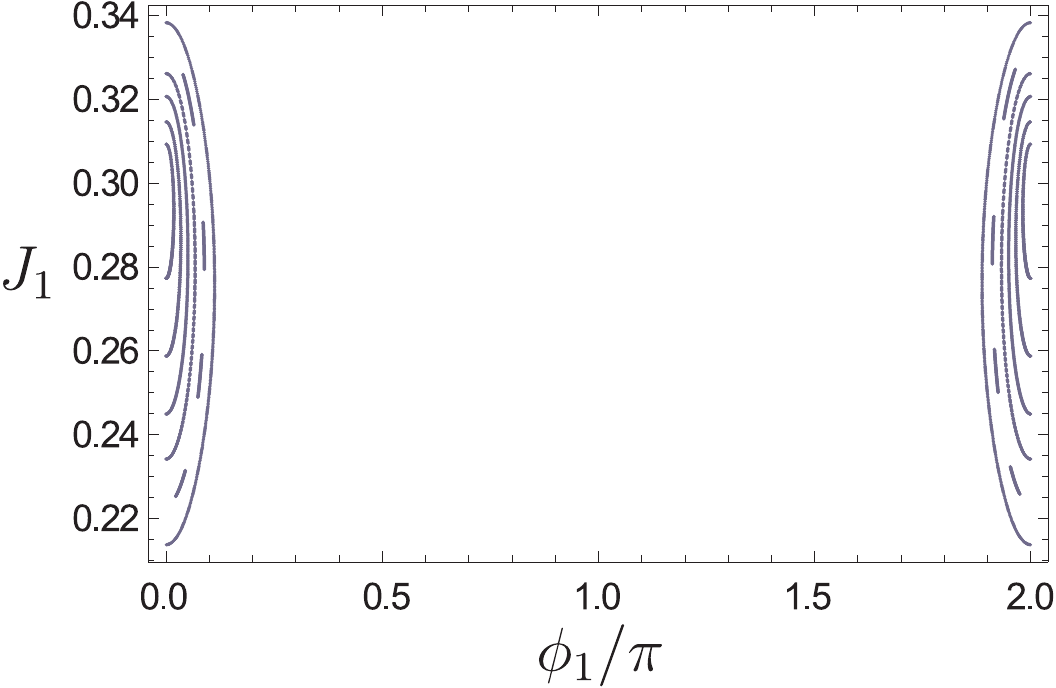}	\\[10pt]
		\end{tabular}
	\caption{(Color online) The Poincar\'e sections are shown for $0.44\lesq L\lesq0.55$}
	\label{fig:pnc_mL2}
\end{figure}

In Fig.~\ref{fig:pnc_mL2} we see how the two asymmetric regions are combined in order to form a single central region. Actually, in the Poincar\'e sections for $L=0.44$ and $L=0.45$ we can see how the asymmetric periodic orbit is generated by the interaction of the invariant curve with the boundary. As the value of $L$ is increased,  both the lower and upper regions of ordered dynamics become extinct for $L=0.47$. In what follows, we can observe, in the sections for $L=0.473$ and $L=0.49$, how the section is separated in two parts. Due to the cyclic nature of the $\phi_1$ variable, rather than being separated, the section 
is concentrated around $\phi_1=0$. Finally, 
as can be seen from the section for $L=0.55$, all trajectories become regular and are confined around the collision orbit in order to present a near integrable picture.

\subsection{Dynamics in different energy levels}\label{last}
The whole study has been focused so far on the value of the energy $h=-0.7475$. This energy has been chosen as a ``characteristic'' one since it corresponds to a typical vortex configuration. By the term typical, we mean here
 a configuration where the vortices are neither too close to each other, 
nor too close to the Thomas-Fermi radius of the condensate. By considering other values of the energy, e.g. in the range $-1.1\leqslant h \leqslant-0.5$, which are physically meaningful, we observed qualitatively the same behavior of the system by varying the value of $L$. For low values of $L$ the system is fully organized having regular orbits. For some value of $L$ the central periodic orbit is getting destabilized through a pitchfork bifurcation, and a chaotic region is created. This region is getting wider as $L$ increases. For even larger values of $L$, the permitted area of the Poincar\'e section shrinks and finally all the permitted configurations of the system correspond to regular orbits which are concentrated around the $S_1 - S_3$ collision orbit. In general we can summarize the behavior of the system by mentioning that the motion of the system is regular when the initial configuration is close to the one or two-vortex regimes which correspond to the integrable cases of having just one or two vortices consisting the system. By one vortex regime here we mean a configuration where the three vortices are far enough from each other so that the interaction between them is weak. On the other hand by two-vortex regime we imply the configuration where two vortices are interacting strongly but are well separated from the third. Finally, when the motion of each vortex is strongly affected by its interaction with both of the other two, then the majority of the orbits are chaotic.

\section{Comparison of the ODE model with PDE computations}

One natural question that arises concerns the validity of our 
ODE model conclusions in connection to the full system. In particular,
while the validity of our particle approach in some regular
region of the system's parameter space may be reasonable
to expect, it is, arguably, a more bold assumption in the regimes
where chaotic dynamics is predicted. It is in that light that we
hereafter present a comparison of our results at the ODE level
with the full Gross-Pitaevskii equation (GPE) model i.e., the corresponding
PDE model.

\begin{figure}
\begin{tabular}{ccc}
\hspace{-1.5cm}\includegraphics[width=6cm]{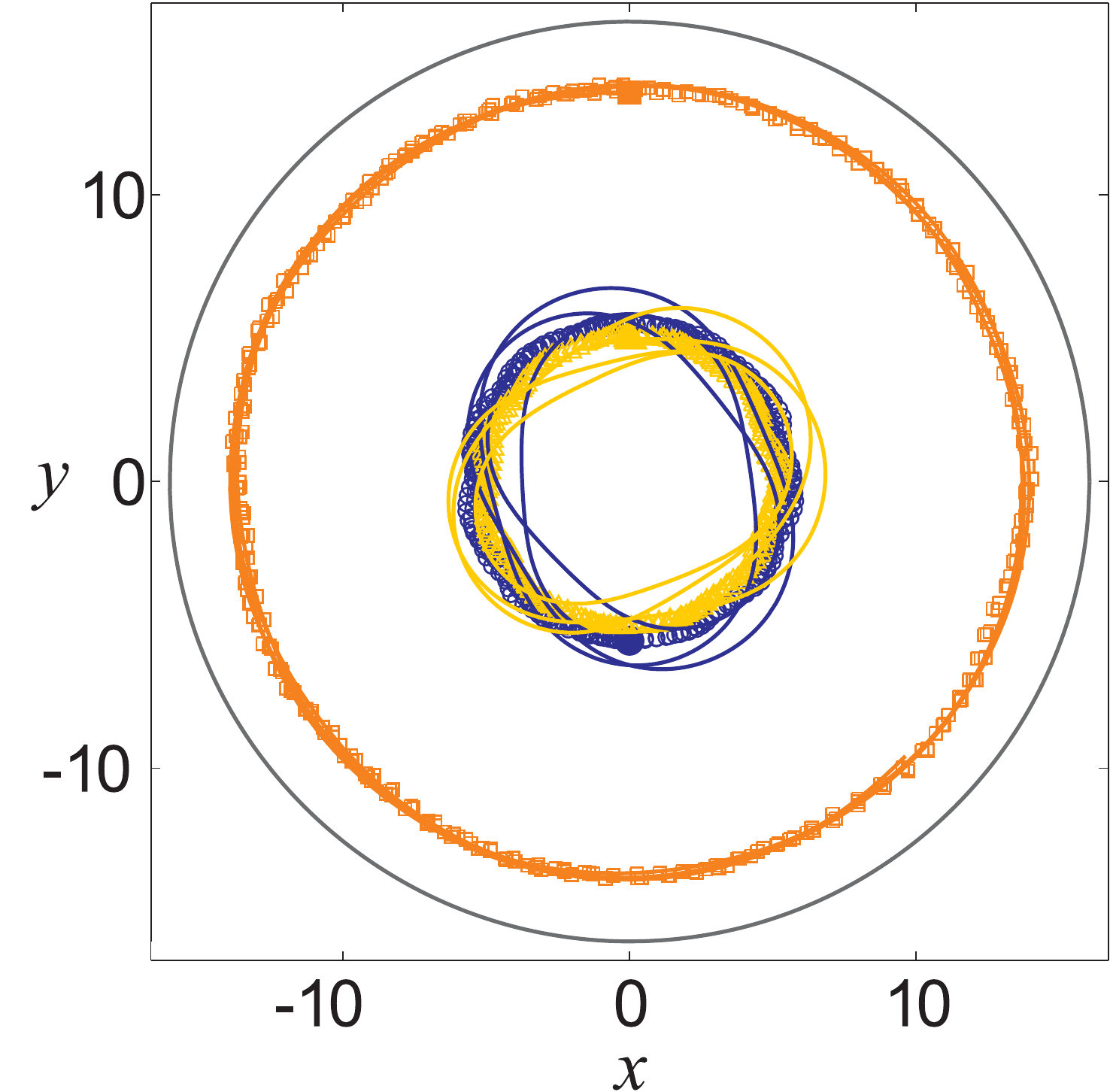} &
\includegraphics[width=6cm]{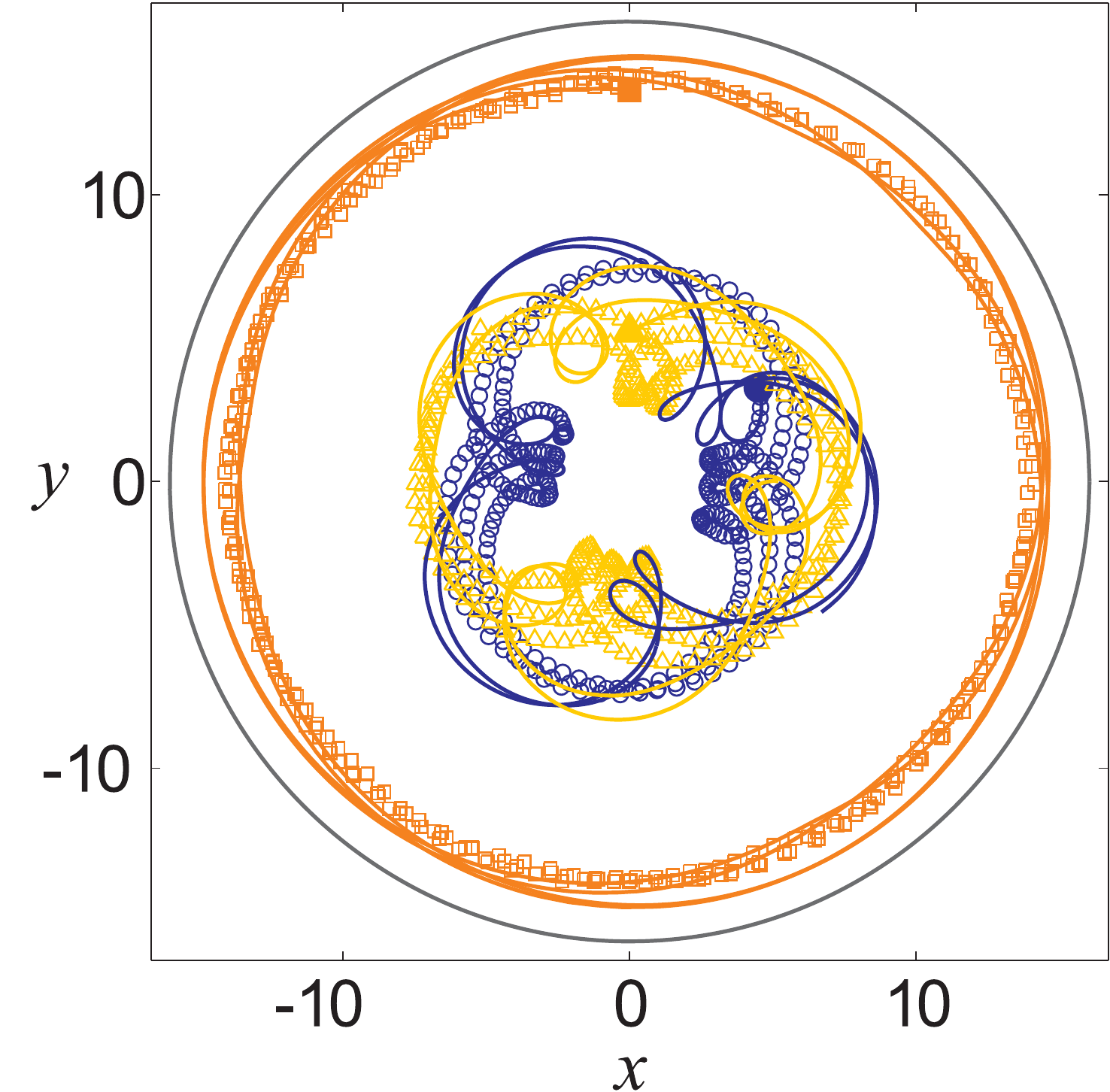} &
\includegraphics[width=6cm]{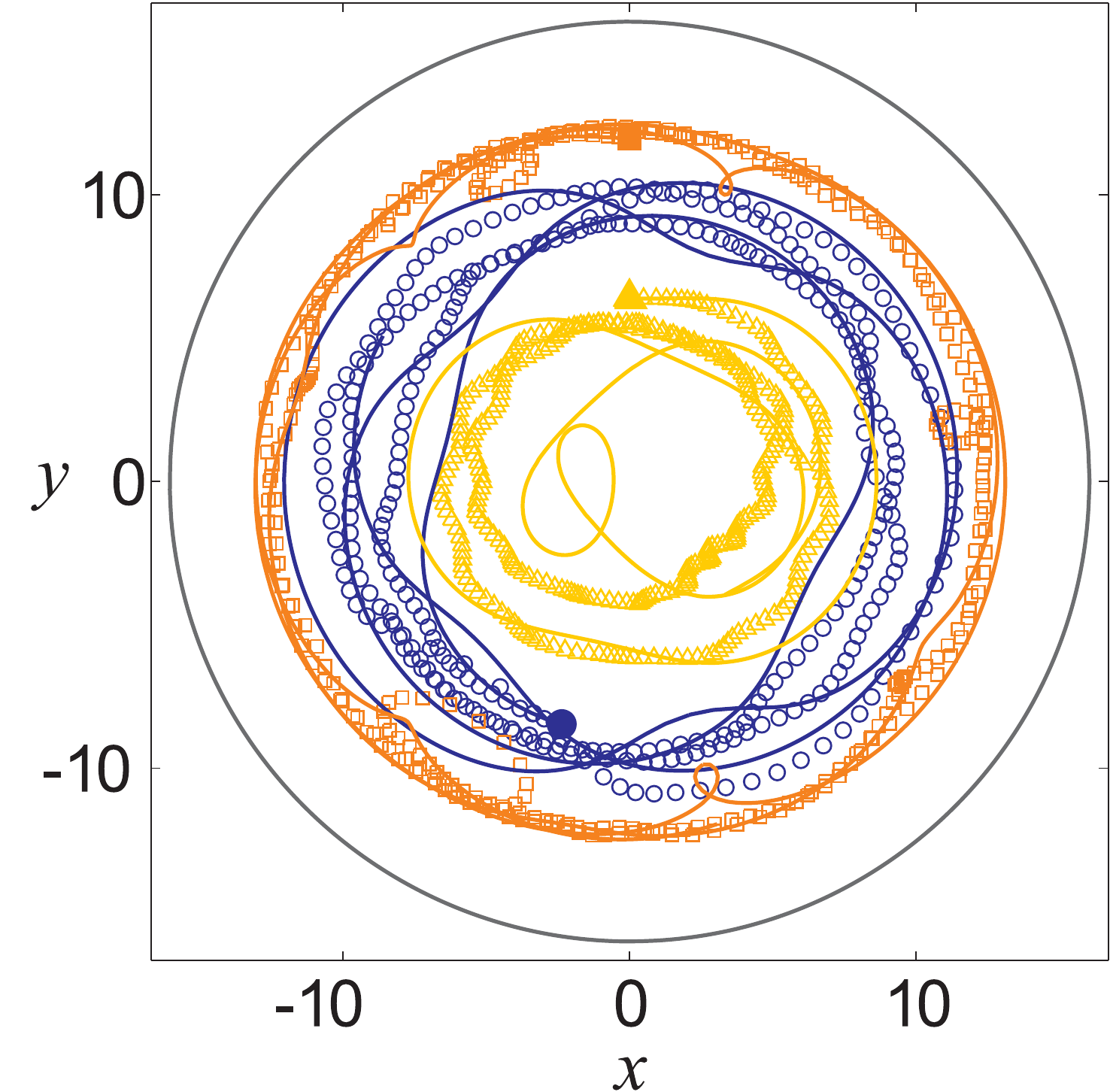} \\
\end{tabular}
\caption{(Color online) Comparison of vortex trajectories in the case of highly regular (top),
less regular but still non-chaotic (middle) and finally in the case of
a weakly chaotic trajectory. The solid lines indicate the ODE results,
while the corresponding symbols the PDE ones. The grey circle denotes
the Thomas-Fermi radius.}
\label{rfig1}
\end{figure}

The results of our comparison are shown in Fig.~\ref{rfig1}. 
In all 3 cases shown (a subset
of a larger number of simulations performed --see also the discussion
below--), consistently the same colors i.e.~yellow, orange and blue (light, intermediate and dark grey) have been used to illustrate
the 3 different vortices. Additionally, in all cases, the solid lines
have been used to denote the ODE results, while the symbols (triangles, squares and
circles) have been used to illustrate the corresponding
PDE ones. Furthermore, the evolution has been given up to $t=500$
in both ODE and PDE and represented in the $x-y$ plane of the dynamics.

At the PDE level, the GPE solved is of the form:
\begin{eqnarray}
i u_t = - \frac{1}{2} \Delta u + V(r) u + (|u|^2-\mu) u
\label{pde1}
\end{eqnarray}
is used. Here, the parabolic trapping potential is of the form
$V(r)=(1/2) \Omega^2 r^2$. The parameters used here are $\mu=16.1$
and $\Omega=0.3538$, which correspond to the experimental setup used 
in~\cite{dsh1,dsh2,dsh3}. It should be noted here that for the comparison of the evolution of (\ref{pde1}) with the ODE model, the original equations (\ref{middlecamp_x}, \ref{middlecamp_y}) have been used. In addition, in order to quantitatively
compare the particle model to the PDE one, a slight technical modification
was used (in comparison to Eqs.~(\ref{eq_mot_norm}), namely 
$\omega _{\mathrm{pr}}(r)=\omega _{\mathrm{pr}^{0}}/(1-\alpha r^2)$ 
with $\alpha=0.78$ was
used as for large distances from the trap center (which some of
our trajectories entailed), this has been found to yield a slightly 
more accurate description of the precession frequency of an isolated vortex 
precessing around the trap center.

The first orbit shown corresponds
to a highly regular trajectory in the case of $L=-0.25$
whereby one of the vortices (the one with negative charge)
is rotating fast in the periphery of the cloud, while
the other two (positively charged ones) rotate close to the
center. The second orbit still pertains to the same angular momentum,
but clearly the relevant trajectory is more complex (yet still regular)
involving a more pronounced quasi-periodic character. Finally, the
third orbit is for $L=-0.05$, in this case existing within the
(weakly) chaotic regime of the dynamics.

A common conclusion from all the cases is that while the 3-degree-of-freedom 
ODE model does not provide a perfect match for its infinite dimensional
PDE counterpart, nevertheless, it provides a rather accurate prediction
of the resulting motion for {\it all} of the above cases. This encompasses
even complex dynamical features such as the meandering present in the
quasi-periodic motion of the second orbit. Notice that in cases like 
the third weakly chaotic orbit of higher angular momentum, the trajectory
may appear slightly less accurate than the earlier, more regular
cases. Nevertheless, it still captures the gross features of the 
dynamics and the apparently (and gradually) space filling nature 
of the chaotic trajectories of the vortices.

There are numerous reasons as to why the agreement may not be 
more quantitatively
accurate. Perhaps the most important one is that indeed we are approximating
the infinite degree of freedom PDE with a 3 degree of freedom, far simpler
dynamical system. Admittedly, we may not capture all of the dynamical features
of the former within 
the latter (especially so in the current Hamiltonian realm).
Weak sound waves can be generated (and are indeed rather unavoidable
in our initialization); these waves create weak interference and
reflection features
that slightly affect the vortex trajectories. Additionally, there are slight
inaccuracies in the tracking of the vortex trajectories, importantly
the vortices are not purely point particles, and finally the screening
effect of the inter-vortex interaction is not fully captured in our
dynamical model (among other things).  Nevertheless, all these issues
considered, and exactly because of their weak nature, the particle model
still provides us with a very accurate qualitative predictor of the dynamics,
remarkably, even when the latter is chaotic (already at the level of the
3 degree of freedom system) and hence presenting a sensitive dependence
on its initial conditions.

Admittedly, this last feature becomes progressively more important, the
stronger the chaoticity of the trajectories. Indeed, for completeness
we should note here that we also considered highly chaotic trajectories
e.g. for $L=0.25$. In some such cases, we observed a substantial departure
of the ODE trajectories from the corresponding PDE ones. However, we
would argue that such a feature is rather reasonable to expect, given
the especially sensitive nature of the corresponding examples to
initial conditions. Broadly speaking, we believe that our finding
that both ordered and even weakly chaotic trajectories can be
adequately followed by the ODE model in a semi-quantitative fashion
for long evolution times such as $t=500$ herein, lends the necessary
credibility to our model and to its findings regarding the existence
of such trajectories and the separation of regular from chaotic
regions, which is one of the principal features of our analysis.

\section{Conclusions}
In this work we have provided a detailed study of a dynamical system which describes the motion of three interacting vortices in a confined Bose-Einstein condensate. The vortices under consideration are non-co-rotating, with 
two of them having positive charge $S_1=S_3=1$ while the remaining having negative charge $S_2=-1$. This system can be studied as a Hamiltonian system of three degrees of freedom, having two integrals of motion; the energy $H$ and the angular momentum $L$. By applying suitable canonical transformations we can bring it to a reduced form of two degrees of freedom with $L$ as a parameter. In 
order to study this system we numerically 
construct a series of Poicar\'e sections for varying $L$.

Our results show that for small values of the angular momentum the system behaves regularly, showing two qualitatively different regions in the phase space. One around a stable periodic orbit and the other around a collision orbit. This means that all the permitted configurations correspond to regular orbits. As $L$ increases, chaotic orbits begin to exist. As the value of $L$ increases further, the region of chaotic motion grows larger but there are always islands of regularity of significant area. As the value of $L$ increases further, the area of permitted orbits of the system on the section decreases. For some value of $L$, it also becomes disconnected and finally it concentrates to a small area around the collision orbit where all the permitted configurations correspond to regular orbits and the system exhibits a behavior close to the one of an integrable one. Although the full study has been presented for a specific value of the energy, we have traced, by performing the same systematic
exploration for other values of the energy, the principal features
that are rather global in this three-vortex system. 

This study may be considered as a starting point for a more detailed
examination of ordered and chaotic features of multi-vortex cluster dynamics
in isotropic (and possibly also anisotropic) 
Bose-Einstein condensates. Generalizing relevant notions
to coherent structures of higher dimensions such as vortex-rings~\cite{konst} 
in three-dimensional Bose-Einstein condensates~\cite{komineas} would
also be a direction of interest for future work. Such studies will
be reported in future publications.

\begin{acknowledgments}
In this research, V.K. and G.V. have been co-financed by the European Union (European Social Fund - ESF) and Greek national funds through the Operational Program "Education and Lifelong Learning" of the National Strategic Reference Framework (NSRF) - Research Funding Program: THALES. Investing in knowledge society through the European Social Fund. V.K.~would also like to thank the research committee of the Aristotle University of Thessaloniki for its support through the postdoctoral research award ``Aristeia ''.

P.G.K. gratefully acknowledges support from US NSF via grants
DMS-0806762, CMMI-1000337, from the Binational Science Foundation 
through grant 2010239, from the US AFOSR through grant 
FA9550-12-1-0332. 

The authors are particularly thankful to
Ricardo Carretero-Gonz{\'a}lez for relaying the relevant parameters,
as well as for technical assistance in connection to the PDE numerical
computations. Finally, the authors would also like to thank Nikos Kyriakopoulos for his valuable remarks.

\end{acknowledgments}

\end{document}